\documentclass[ amsmath,amssymb,aps,prl,reprint]{revtex4-2}

\usepackage{graphicx}
\usepackage{dcolumn}
\usepackage{bm}
\usepackage{float} 
\usepackage{hyperref}
\hypersetup{colorlinks,urlcolor=blue,citecolor=black,linkcolor=black}

\begin{document}

\title{\bfseries Long-lived magnetization in an atomic spin chain tuned to a diabolic point}
\author{R.J.G. Elbertse$^{\rm 1}$, D. Borodin$^{\rm 2}$, J. Oh$^{\rm 2,3}$, T. Ahn$^{\rm 2,3}$, J. Hwang$^{\rm 2,3}$, J.C. Rietveld$^{\rm 1}$, \\ A.J. Heinrich$^{\rm 2,3}$, F. Delgado$^{\rm 4}$$^*$, S. Otte$^{\rm 1}$$^*$, Y. Bae$^{\rm 2,3}$$^*$$^\dagger$ \\
\textit{\small{$^{\rm 1}$Department of Quantum Nanoscience, Kavli Institute of Nanoscience,} \\
\small{Delft University of Technology, Delft, The Netherlands} \\ 
\small{$^{\rm 2} $Center for Quantum Nanoscience, Institute for Basic Science (IBS), Seoul, South Korea} \\ \small{$^{\rm 3} $Department of Physics, EWHA Womans University, Seoul, South Korea} \\ 
\small{$^{\rm 4} $Instituto Universitario de Estudios Avanzados IUDEA, Departamento de F\'{i}sica, Universidad de La Laguna} \\
\small{La Laguna, Tenerife, Spain} \\
\small{$^\dagger$Current address: Empa, Swiss Federal Laboratories for Materials Science and Technology,} \\ \small{nanotech@surfaces Laboratory, Switzerland}
}}

\date{\today}
\begin{abstract}
Scaling magnets down to where quantum size effects become prominent triggers quantum tunneling of magnetization (QTM), profoundly influencing magnetization dynamics. Measuring magnetization switching in an Fe atomic chain under a carefully tuned transverse magnetic field, we observe a non-monotonic variation of magnetization lifetimes around a level crossing, known as the diabolic point (DP). Near DPs, local environment effects causing QTM are efficiently suppressed, enhancing lifetimes by three orders of magnitude. Adjusting interatomic interactions further facilitates multiple DPs. Our study provides a deeper understanding of quantum dynamics near DPs and enhances our ability to engineer a quantum magnet.
\end{abstract}

\maketitle

\begin{figure}
    \centering

    \includegraphics[width=\linewidth]{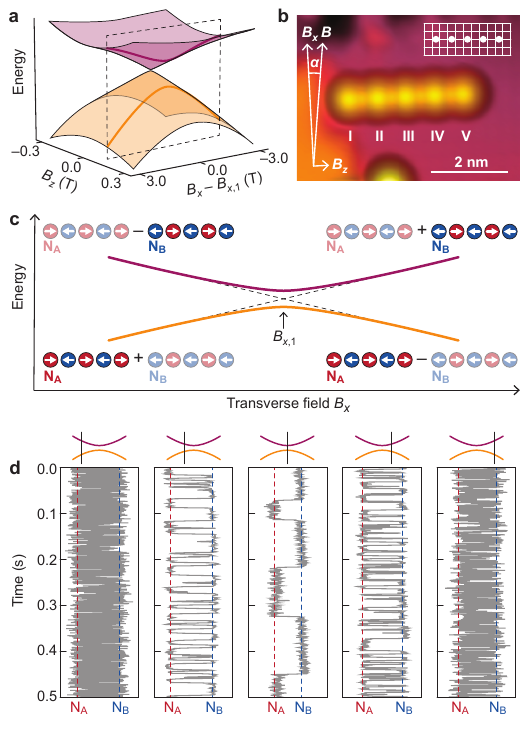}
    \caption{Diabolic points in Fe atomic chains on Cu$_2$N. (a) Energy levels for the two lowest states of a single Fe atom (orange for the ground state $\psi_0$ and pink for the first-excited state $\psi_1$) with a diabolic point at the crossing. Thick lines show a cut of the energy surfaces, indicating the corresponding energies of the two states at a finite $B_z$ outlined by the dashed rectangle. (b) Constant-current STM image of a Fe$_5$ chain ($V_\mathrm{DC} = 100$~mV, $I=10$~pA, $B_x = 2$~T). Atoms are labeled in Roman numerals (I--V). Schematics shows Fe atoms (white circles) on top of Cu$_2$N lattice. The intersections of grid lines correspond to the nitrogen atom position in Cu$_2$N. Magnetic field directions are tilted by the angle $\alpha$ with respect to the crystal axis. (c) Schematic overview of state composition for $\psi_0$ and $\psi_1$ of Fe$_5$ along a similar cut as in panel a. (d) Current traces taken for atom III of antiferomagnetically coupled Fe$_5$ and different magnetic fields around the DP, located at $4.2$~T, indicated in the schematics above. Unless stated otherwise, all traces are obtained at 3~mV and approximately 10~pA, where magnetization switching occurs due to QTM and over-the-barrier excitation can be neglected, see Supplementary Note 8 for further details.}
    \label{fig:1}
\end{figure}

In quantum mechanical systems, unusual dynamic processes occur when energy levels approach and mix with each other. In a two-parameter space, the degeneracy between orthogonal states creates a level crossing of energy surfaces (Fig.~\ref{fig:1}a), the shape of which reminds of the toy, \textit{diabolo}, and thus is dubbed a diabolic point (DP) \cite{BerryProcRS1984}. This DP has attracted significant attention in quantum magnets \cite{WernsdorferScience1999}, which are characterized by two metastable magnetization states separated by an energy barrier \cite{Sessoli1993, Thomas1996}. In the vicinity of the DP, quantum tunneling of magnetization (QTM) between these states is suppressed due to destructive interference among separate tunneling paths \cite{vonDelftAPS1992, LossPRL1992,Garg_epl_1993}. While the importance of  DPs has been shown from ensembles of molecular magnets \cite{WernsdorferScience1999,BurzuriPRL2013,WernsdorferPRL2005}, precise control of local environments as a control knob of DPs has remained elusive.

Manipulation of magnetic atoms with a scanning tunneling microscope (STM) allows for the assembly of prototypical quantum magnets with controllable energy barriers ranging from 100~$\mu$eV to 100~meV \cite{HirjibehedinScience2007,NattererPRL2018,Singha2021}. The lifetime of magnetization states in these magnets can be determined by monitoring the spin polarized current through one of the magnet's atoms over time \cite{LothScience2012}. When the magnetic anisotropy barrier exceeds the thermal energy, the lifetime is dominated by through-the-barrier transitions, i.e. QTM, resulting from hybridization between quantum states on either side of the barrier. While various systems have been studied with different degree of spin state hybridization \cite{LothScience2012} and spin-spin interactions \cite{LothScience2012,SpinelliNatureMaterials2014}, it remains challenging to vary individual parameters due to the discrete nature of binding sites on surfaces. Instead, a more effective control knob for QTM may be achieved by exploiting the physics of a DP, where the hybridization of the quantum states is expected to quench, allowing, at least in principle, arbitrarily long lifetimes.

In this work, we demonstrate the manifestation of DPs through the spin dynamics of nanomagnets by assembling Fe atoms into chains on Cu$_2$N/Cu(100). Precisely adjusting chain length and interatomic spacing enables us to tailor the spin-spin interactions and thereby to engineer DPs in a controlled manner. When tuning the direction and strength of the external magnetic field near a DP, we observed a significant increase of magnetic lifetimes, with enhancement of up to three orders of magnitude. We provide a comprehensive picture of the quantum state composition near the DP, offering a rational strategy to control the spin dynamics of quantum magnets.

To model a chain of $N$ Fe atoms on Cu-sites of Cu$_2$N, we consider a spin Hamiltonian that includes the Zeeman energy, the uniaxial and transverse magnetic anisotropy terms for each atom, as well as the Heisenberg exchange interaction between neighboring atoms \cite{LothScience2012,HirjibehedinScience2006,YanNature2014,ElbertseCommPhysics2020}:
\begin{equation}\label{eq:Hamiltonian}
\begin{split}
    H = & \sum_i^N \left[g_i\mu_{\mathrm B} \boldsymbol{B}^\mathrm{tot}_i\cdot \boldsymbol{S}_i + D_iS_{i,z}^2 + E_i(S_{i,x}^2 - S_{i,y}^2)\right] \\ 
    & + \sum_i^{N-1} J_{i}\boldsymbol{S}_i\cdot\boldsymbol{S}_{i+1}.
\end{split}
\end{equation}
For an Fe atom on site $i$, $\boldsymbol{B}^\mathrm{tot}_i = \boldsymbol{B} + \boldsymbol{B}^\mathrm{tip}_i$ is the total magnetic field composed of external and tip fields, $\mu_\mathrm{B}$ the Bohr magneton, and $\boldsymbol{S}_i$ the spin operator with a magnitude of $S_i=2$. Note that we allow for subtle variations in the values of the g-factor $g_i$, anisotropy parameters $D_i$ and $E_i$, and exchange interaction strength $J_{i}$ between atoms in the chain, since these parameters might vary due to subtle changes in the local strain in the underlying Cu$_2$N layer, as evidenced by variations of Hamiltonian parameters throughout the literature (see Table~S1). The easy-axis $z$ is oriented along the in-plane Cu-N bonds \cite{HirjibehedinScience2007}; we define the $x$ and $y$ axes as the remaining in-plane direction and the out-of-plane direction, respectively (Fig.~\ref{fig:1}b). Since the chain is not perfectly aligned with the external magnetic field in the experiment, the angle $\alpha$ is used to decompose the external field into both transverse ($B_x$) and longitudinal ($B_z$) components. 

We investigate the DPs as a function of the transverse magnetic field, $B_x$. Solving the spin Hamiltonian (Eq.~\ref{eq:Hamiltonian}) for a single Fe atom ($N=1$) gives the analytical solution for the $B_{x,n}$ fields where the DPs appear \cite{BrunoPRL2006,ZitkoNJP2010}:
\begin{equation}
    B_{x,n} = \frac{n\sqrt{2E(E-D)}}{g \mu_\mathrm{B}}.
\end{equation}
Here, $n$ is the diabolic point index that ranges from  $2S-1$ to $1 - 2S$ in double-integer steps. To estimate the location of DPs, we use Hamiltonian parameters which have been obtained from previous work, see Supplementary Note 1. For a single Fe atom at the Cu site, the lowest positive magnetic field DP ($n = 1$) is expected at $B_{x,1} \approx 9.5$~T. A second DP ($n = 3$) can be reached at even larger values of $B_x$. As depicted in Fig.~\ref{fig:1}a, when $B_x = B_{x,1}$ and $B_z = 0$, an energy level crossing occurs between the two lowest-lying eigenstates, $\psi_0$ and $\psi_1$. When a small $B_z$ is applied, sweeping $B_x$ results in an avoided level crossing, as shown by the surface cut in Fig.~\ref{fig:1}a. 

Diabolic points in atomic chains of length $N > 1$ with $|J/D| \lesssim 1$ can be understood in a similar fashion, although there are now $N$ diabolic points for each diabolic point index, leading to a total of $2NS$ diabolic points. Now, the spin-spin interaction between the atoms can be used as a control knob for the location of DPs as a function of $B_x$. By adjusting the number of atoms in the chain and the interatomic distance, we are able to precisely determine the magnetic field values at which DPs are expected to occur (see Supplementary Note 2). 

The magnetic fields needed to identify the DP in a single Fe atom are beyond available transverse magnetic field ($6$~T) of our instrument \cite{HirjibehedinScience2007,ZitkoNJP2010}. However, for longer chains, the DP eventually becomes accessible within the range of our experimental capabilities. Our initial calculations using Eq.~\ref{eq:Hamiltonian} predicted a DP for an antiferromagnetically coupled Fe$_5$ chain (Fig.~\ref{fig:1}b) at $B_{x,1} \sim 4$~T (see Supplementary Note 2), which guided our spin lifetime measurements. We investigate the influence of DPs on the magnetization bistability of the Fe atomic chains with spin-polarized STM.  

Figure~\ref{fig:1}c schematically shows the energy levels of the two lowest-lying states ($\psi_0$ and $\psi_1$) in the antiferomagnetic Fe$_5$ chain as a function of transverse magnetic fields. For $B_x \ll B_{x,1}$, these two states are mainly composed of N\'eel states, denoted as $\mathrm{N_A} = \{-2, +2, -2, +2, -2\}$ and $\mathrm{N_B} = \{+2, -2, +2, -2, +2\}$ (expressed in the $S_z$-basis), with subtle contributions from other spin states. Owing to the presence of a finite longitudinal component of the field and $N$ being an odd number, the ground state $\psi_0$ has a larger contribution from $\mathrm{N_A}$ ($>98\%$) compared to $\mathrm{N_B}$, while the opposite holds for $\psi_1$. The finite contributions of both $\mathrm{N_A}$ and $\mathrm{N_B}$ in these two states demonstrate that the N\'eel states are hybridized, enabling QTM between them. 

In $\psi_0$, the contributions of $\mathrm{N_A}$ and $\mathrm{N_B}$ are symmetric, whereas in $\psi_1$ they are antisymmetric. Around $B_x=B_{x,1}$, the two states undergo an avoided level crossing, beyond which their symmetry is inverted. At the DP, the contribution of the minority N\'eel state vanishes, significantly enhancing the purity of $\psi_0$ and $\psi_1$ as mainly composed of $\mathrm{N_A}$ and $\mathrm{N_B}$, respectively, thereby suppressing QTM. 

Using spin-polarized STM, we are able to capture the time-dependent magnetization switching of the Fe chain at different $B_x$. By positioning the tip above one of the Fe atoms in the chain, we observe telegraph noise in the current signals, arising from the magnetization switching between the two lowest-lying states of the chain (Fig.~\ref{fig:1}d). The specific spin polarization of the tip, and which atom in the chain is being probed, determines the current value characteristic for $\mathrm{N_A}$ and $\mathrm{N_B}$. For $B_x \ll B_{x,1}$, we detect rapid, yet clearly distinguishable switching events between two distinct current values. As $B_x$ approaches the DP, the switching rate markedly decreases, to gradually increase again beyond the DP. 

\begin{figure}[H]
    \centering
    \includegraphics[width=\linewidth]{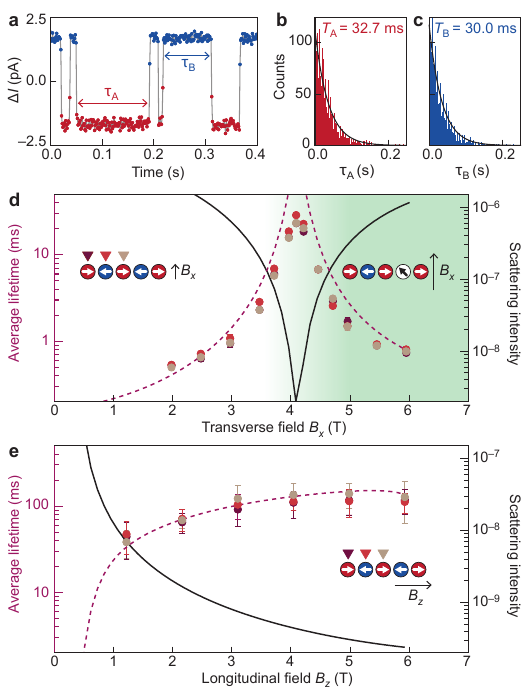}
    \caption{Lifetime of magnetization states. (a) A current trace obtained at a constant tip height on atom III of the Fe$_5$ chain near the diabolic point ($B_x = 4$~T). A spin-polarized tip reveals two magnetization states of the chain: $\mathrm{N_A}$ (red) and $\mathrm{N_B}$ (blue). The data was shifted by about 8~pA to center the middle values around zero, mitigating background fluctuations caused by drift. (b,c) Histograms of lifetimes $\tau_\mathrm{A}$ and $\tau_\mathrm{B}$, as defined in a, with a fit to an exponential function to determine lifetimes $T_\mathrm{A}$ and $T_\mathrm{B}$, respectively. (d) Average lifetimes $T_{\rm avg}$ measured for atoms I, II and III of an antiferromagnetic Fe$_5$ chain at different $B_x$. The corresponding longitudinal component $B_z$ is approximately $0.35\%$ of the transverse magnetic field, due to a small angle ($\alpha \approx 0.2^{\circ}$) between the external magnetic field and the crystal axes. Error bars are the standard deviation $\sigma$. The lifetime and the scattering intensities calculated using master rate equations are given in the purple dashed and black solid lines, respectively. Background color and diagrams indicate the quanta of $S_x$ in the ground state: zero before the diabolic point and one after the diabolic point, with the diabolic point at $B_{x,1} \approx 4.1$~T. Different colors of data points correspond to measurements performed on different atoms along the chain, see diagram. (e) Lifetime obtained similar to d, but for a magnetic field applied along the longitudinal axis. }
    \label{fig:2}
\end{figure}

To quantitatively investigate the evolution of spin dynamics around the DPs, we extract the values of spin lifetimes from a current trace as demonstrated in Fig.~\ref{fig:2}a. Shown are the lifetimes $\tau_\mathrm{A}$ and $\tau_\mathrm{B}$, representing the duration between consecutive switches in eigenstates dominated by $\mathrm{N_A}$ and $\mathrm{N_B}$, respectively. By collecting sufficiently long traces, we obtain histograms for $\tau_\mathrm{A}$ (Fig.~\ref{fig:2}b) and $\tau_\mathrm{B}$ (Fig.~\ref{fig:2}c), enabling us to extract characteristic lifetimes $T_\mathrm{A}$ and $T_\mathrm{B}$, respectively \cite{LothScience2012}. Finally, we define $T_{\rm avg} = \left(T_{\mathrm{A}}^{-1} + T_{\mathrm{B}}^{-1}\right)^{-1}$ as the average lifetime of the magnetization states. The extracted $T_{\rm avg}$ for atom I--III of the antiferromagnetic Fe$_5$ chain are shown as a function of the transverse magnetic field in Fig.~\ref{fig:2}d, showing a pronounced peak at $B_x = 4.1$~T spanning two orders of magnitude. As we will demonstrate below, we can associate this field value to the first DP of the chain $B_{x,1}$. This DP coincides with a minimum in the scattering amplitude, shown by the black solid line in Fig.~\ref{fig:2}d and defined as $\sum_{a=x,y,z}|\langle \psi_0|S_{a,i}|\psi_1\rangle|^2$, which is an indication of the hybridization between $\psi_0$ and $\psi_1$. 

Note that the presence of a small longitudinal field induces an avoided level crossing instead of a level crossing, see Supplementary Notes 3 and 6 for further details. This provides the peak in the lifetime with a finite width, enabling us to measure it. The observed trend happens for all atoms of the chain, and is robust to many experimental parameters (see Supplementary Note 4).

The dashed line is a simulation of the lifetime measurements of the chain based on master rate equations (see Supplementary Note 2). Neither the simulations nor the measurements show a peak reaching infinity, owing to minute contributions of states other than $\mathrm{N_A}$ and $\mathrm{N_B}$. The experimental data shows slightly lower lifetimes than the simulation, especially close to the diabolic point. We attribute this deviation to accidental high-energy electrons caused by voltage noise, leading to over-the-barrier excitations. 

Until now we have described the situation in terms of the $S_z$-basis. However, for interpretation purposes, it is insightful to consider the situation in terms of the $S_x$-basis. For $B_x < 4.1$~T, the expectation value of $S_x$ in the ground state approaches zero, indicated by the white background on the left-side of Fig.~\ref{fig:2}d and the diagram on the left hand-side. The first excited state contains a single quantum of $S_x$ (i.e. $| \langle S_x\rangle |= 1$). Past the DP, i.e. the avoided level crossing, the states are inverted, transferring the finite $S_x$ magnetization to the ground state, as indicated by the green color of the right-side of Fig.~\ref{fig:2}d and the diagram on the right hand-side. Note that the ground state now consists of a superposition of five spin states, each having the quantum of $S_x$ on a different atom. 

We also demonstrate that the increase of lifetime through the DP emerges only for specific orientation of the magnetic field with respect to the quantization axis. When the magnetic field is swept along the easy axis of the antiferromagnetically coupled Fe$_5$ chain, no peak in the lifetime is observed, see Fig.~\ref{fig:2}e. The lifetime slightly increases as a result of decreasing scattering intensity. Within the range of available magnetic fields of $B_z$ (up to $\sim 6$~T), there is no energy level crossing. Thus, this behaviour is strictly monotonic and results from an increasing imbalance in the N\'eel state contributions in each eigenstate. 

\begin{figure}
    \centering
    \includegraphics[width=\linewidth]{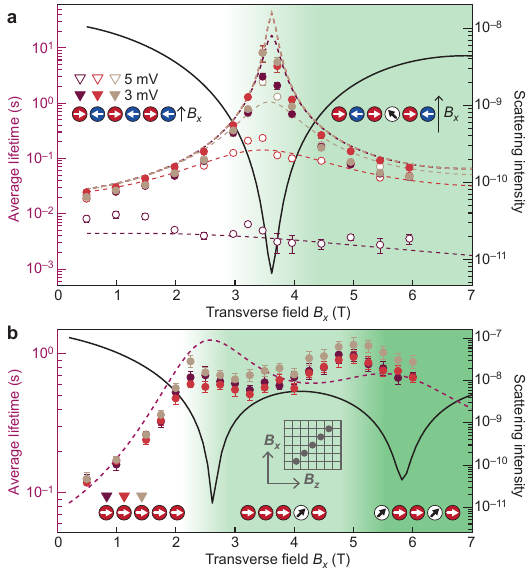}
    \caption{Tuning diabolic points. (a) Lifetime of an antiferromagnetic Fe$_6$ chain. Data points given in filled and empty circles were obtained at $3$~mV and $5$~mV, respectively, for atoms I-III in the chain. Dashed lines are from a simulation of the experiment. Background color and diagram indicate the quanta of $S_x$ in the ground state, which increases from zero to one upon passing the diabolic point. The solid black line indicates the scattering amplitude from the simulation. In this measurement,at $10$~pA, $1.3$~K, $\alpha \approx 5^{\circ}$. (b) Similar to panel a, but for a ferromagnetic Fe$_5$ chain, and $\alpha \approx 0.2^{\circ}$. Background color and diagrams indicate a second quantum of $S_x$ entering the ground state as the second diabolic point at around $5$~T is passed. Note that the easy axis $z$ and hard axis $x$ are defined along the crystallographic directions, see inset. Error bars indicate 2$\sigma$.}
    \label{fig:3}
\end{figure}

The finite energy difference between the two lowest energy states at the avoided level crossing is responsible for the width of the observed peak in Fig.~\ref{fig:2}d and therefore limits the efficiency of increasing spin lifetimes. The energy difference at the DP emerges as a consequence of Zeeman energy, resulting from a small angle $\alpha$ between the quantization axis of the Fe atoms with respect to the applied magnetic field. This suggests that in order to achieve a sharper peak, one must achieve a null magnetic field along $B_z$. This is practically impossible, as $\alpha$ will inevitably be nonzero in our experimental setup. An alternative approach would be to make use of even-length antiferromagnetic chains, where both N\'eel states have equal energy, irrespective of $B_z$. In line with this idea, Fig.~\ref{fig:3}a shows spin lifetimes measured on a Fe$_6$ chain, where a sharper peak than on the Fe$_5$ chain is observed.

Despite the expected absence of Zeeman splitting between the two states, the observed peak shows a finite width, which implies that there must be factors breaking the symmetry between the N\'eel states. Our analysis (see Supplementary Note 5) suggests that the primary cause of the asymmetry is variations in the $g$-factors of the atoms in the chain. The asymmetry introduced from the tip field has a minor importance. This difference in $g$-factors results in an energy discrepancy of approximately 50 $\mu$eV at $B_1 = 6$~T, favoring one N\'eel state at higher magnetic fields. 

The lifetime of the Fe$_6$ chain increases by nearly three orders of magnitude at the peak, which is shifted to a lower magnetic field compared to Fe$_5$. The overall lifetime has also increased, as larger chains are naturally more stable~\cite{LothScience2012}. It is noteworthy that this considerable enhancement of lifetime near the DP is only apparent when the magnetization lifetime is primarily determined by the QTM. Once the over-the-barrier transitions become frequent, the magnetization anomaly diminishes, and we only observe a subtle change in lifetime, as shown in the lifetime curves measured at 5~mV (see also Supplementary Notes 4, 8 and 9). 
While dehybridization also occurs among higher energy states, as shown in Supplementary Note 2, it appears not to affect the rate of over-the-barrier excitations, for voltages close to the over-the-barrier threshold. 

Antiferromagnetic Fe atom chains showed one DP in the transverse magnetic field ranging from 0~T to 6~T. However, adjusting magnetic interactions between atoms in the chain through atom manipulation gives us the possibility to change the location and spacing of DPs. Our simulations indicate that a ferromagnetic Fe$_5$ chain ($J<0$) is the most likely candidate where multiple diabolic points can be observed, given the magnetic field range of our experimental setup. In this case, the chain tends to behave as a macrospin with a large total spin. When the coupling is sufficiently strong, this leads to $NS$ positive DPs all equally spaced across the transverse magnetic field. The highest-field DP corresponds to the $B_{x,3}$ value of a single Fe, while the lowest-field DP occurs at $B_{x,3}/(2NS-1)$. Consequently, the DPs for ferromagnetic chains occur at lower magnetic field values. 

We built a ferromagnetic Fe$_5$ chain by placing the 5 atoms diagonally with respect to the easy axis, see inset of Figure 3b, leading to $J = -0.7$~meV \cite{SpinelliNatureMaterials2014}. Figure~\ref{fig:3}b shows that this chain exhibits two distinct peaks in lifetimes as a function of $B_x$ within our operation range; one at $2.5$~T and one at $5.0$~T. The second DP is associated with delocalized states where two different atoms in the chain gain a quantum of $S_x$ (i.e. $| \langle S_x\rangle |= 1 \rightarrow 2$). Hence, this may be associated with two-magnon states \cite{BetheZTdMetalle1931,DelgadoArXiv2023}. 

The simulation results, see the dashed line in Fig.~\ref{fig:3}b, qualitatively reproduce the experimental observations but fail to reproduce the exact positions of the DPs. While the scattering intensity does still contain two clearly distinguishable peaks, both the measured and simulated lifetime curves show comparably shallow peaks. We believe that this is due to the overlapping of the two lifetime peaks as well as a larger Zeeman energy associated with the ferromagnetic coupling. The larger energy not only broadens the peaks but also decreases the energy difference between $\psi_1$ and higher energy states, resulting in more over-the-barrier transitions.


Our work presents a comprehensive approach, combining experimental and theoretical methods, to elucidate the physics of magnetization stability of individual atomic spin chains near a diabolic point. We have demonstrated that the switching rate between the two lowest energy levels can be precisely controlled through tailored transverse magnetic fields, which suppresses quantum tunneling of magnetization. This suppression emerges as a consequence of dehybritization of the lowest lying spin states near an avoided level crossing, which, in the case of Fe chains on Cu$_2$N, results in strong enhancement of lifetimes up to three orders of magnitude. 

While effects of diabolic points have been observed previously in single molecule magnets \cite{WernsdorferScience1999,BurzuriPRL2013,WernsdorferPRL2005} through ensemble measurements, we showed that diabolic points in quantum magnets can be manipulated and rationally designed by tailoring the interaction of individual atomic spins on surfaces. The composite nature of atomic spin chains provides the possibility to understand the topology of diabolic points, with different predicted behaviour for ferromagnetic and antiferromagnetic chains, and a crucial role of parity predicted in the latter case. 

The dramatic enhancement of the lifetime provides an interesting avenue into spintronics \cite{ElbertseCommPhysics2020,WernsdorferIntJNanotech2010} and applications of coherent spin dynamics \cite{BaumannScience2015,YangScience2019}. The extreme sensitivity of spin lifetimes of a quantum magnet near a diabolic point could be exploited for precise sensing of local and external magnetic fields at the atomic scale. This sensitivity near diabolic points can also be used to determine parameters of spin Hamiltonians, such a magnetic anisotropy and g-factors, with exquisite precision, leading to further understanding of magnetic material in various applications \cite{MarcanoACS2022,PankhurstApplMNB2003,WernsdorferPRB2004}. 

\vspace{3mm}

R.J.G.E., J.C.R. and S.O. acknowledge support from the Netherlands Organisation for Scientific Research (NWO) and from the European Research Council (ERC Starting Grant 676895 “SPINCAD”). D.B., J.O., T.A., J.H., A.J.H., and Y.B. acknowledge support from the Institute for Basic Science (IBS-R027-D1). Y.B. acknowledges support from Asian Office of Aerospace Research and Development (FA2386-20-1-4052). D.B. acknowledges support from the Alexander von Humboldt Foundation for financial support through a Feodor-Lynen Research Fellowship. F.D. acknowledges support from MCIN/AEI/10.13039/501100011033, and ``FEDER, a way to make Europe", by the European Union (PID2022-138269NB-I00).

\vspace{3mm}
All simulations, raw data, code to process the data, figures and data points on the figures in the main text and the Supplementary Information \cite{supp} are available from the Open Data folder accessible through the digital object identifier ("DOI") \mbox{10.5281/zenodo.10906000}. 

\vspace{3mm}

\nocite{Rossier_prl_2009,Delgado_Palacios_prl_2010,LothScience2010,HirjibehedinScience2007,LothNJP2010,NicklasJApplPhys2011,LothScience2012,BryantPRL2013,SpinelliNatureMaterials2014,YanNanoLetters2015,YanScienceadvances2017,Rolf-PissarczykPRL2017,RudowiczApplMagnRes2019,ElbertseCommPhysics2020,Rossier_prl_2009,Delgado_Palacios_prl_2010,LothNaturePhysics2010,DelgadoProgSS2017,HirjibehedinScience2006,HirjibehedinScience2007,HwangRSI2022,HirjibehedinScience2007,SpinelliNatureMaterials2014,PaulNPhysics2017,YanNature2014,YamadaJSSN2023}

\vspace{3mm}
The authors declare no competing interests.

\vspace{3mm}

\small{$^*$Corresponding authors: F.D. (fdelgadoa@ull.edu.es), S.O. (a.f.otte@tudelft.nl), Y.B. (bae.yujeong@qns.science)} \\

\bibliographystyle{apsrev4-2}

\newpage

\onecolumngrid

\section*{Supplementary Note 1: Parameters used in the simulations}
\vspace{5mm}
\addcontentsline{toc}{section}{Supplementary Note 1: Parameters used in the simulations}

In the Hamiltonian given in the main text (Eq.~1), the parameters ($D_i$, $E_i$, $g_i$, and $J_{ij}$) specified for Fe on Cu$_2$N have been experimentally investigated. The values found in the literature are summarized in Table \ref{tab:OverviewOfParameters}, showing significant variations among adatoms. These discrepancies can be attributed to variations of local environments influenced by factors such as strains in the underlying Cu$_2$N layer, nearby defects, and adjacent atoms. While previous studies derived parameters from bias spectroscopy, our approach involved extracting values only based on lifetime curves, but referred to the literature values in Table \ref{tab:OverviewOfParameters} to facilitate optimizing the parameters for simulations. The simulations are based on the Master Rate Equations~\cite{Rossier_prl_2009,Delgado_Palacios_prl_2010,LothScience2010}. \\

\begin{table}[H]
    \centering
    \begin{tabular}{ccccccc}
        $N$ & $D$ (meV) & $E$ (meV) & $J$ (meV) & $g$ & Source & Year  \\
        \hline
        \hline
        $1$   & $-1.55 \pm 0.01$       & $0.31 \pm 0.01$    &                & $2.11 \pm 0.05$ & \cite{HirjibehedinScience2007}    & 2007\\
        $1$   & $-1.55 \pm 0.02$       & $0.31 \pm 0.01$    &                & $2.11$          & \cite{LothNJP2010}                & 2010\\
        $3$ (sim.)  & $-1.53$          & $0.37$    & $>25$          &                 & \cite{NicklasJApplPhys2011}       & 2011\\
        $\geq3$  &                        &                    & $1.3 \pm 0.1$  &                 & \cite{LothScience2012}            & 2012\\
        $2$   & $-1.87$                & $0.31$             & $0.7$          &                 & \cite{BryantPRL2013}              & 2013\\
        $2$ (FM)& $-1.37$              & $0.31$             & $-0.69$        &                 & \cite{BryantPRL2013}              & 2013\\
        $6$ (FM)& $-1.29$              & $0.31$             & $-0.73$        &                 & \cite{SpinelliNatureMaterials2014} & 2014\\
        $1$   & $-1.59 \pm 0.03$       & $0.31 \pm 0.01$    &                & $2.07 \pm 0.19$ & \cite{YanNanoLetters2015}         & 2015\\
        $3$   & $\{-2.1,-3.6,-2.1\}$   & $0.31$             & $1.15 \pm 0.1$ & $2.1$           & \cite{YanNature2014}          & 2015\\
        $3$   & $\{-2.1,-3.6,-2.1\}$   & $0.3$              & $1.15$         & $2$             & \cite{YanScienceadvances2017}     & 2017\\
        $3$   & $\{-1.88,-2.42,-1.87\}$& $\{0.39,0.2,0.31\}$& $\{0.71,0.66\}$&                 & \cite{Rolf-PissarczykPRL2017}     & 2017\\
        $1$ (sim.)& $-1.53$                & $<0$               &                & $2.14 \pm 0.12$ & \cite{RudowiczApplMagnRes2019}    & 2019\\
        $3,5$ & $-1.87$                & $0.31$             & $0.7$          & $2.11$          & \cite{ElbertseCommPhysics2020}    & 2020\\
        
    \end{tabular}
    \caption{\textbf{Overview of literature values for parameters in the Hamiltonian.} The first column shows the number of atoms $N$. The values given in the table are mostly obtained from experiments conducted on antiferromagnetically coupled Fe atomic chains, except for instances labeled as "(sim.)" and "(FM)", denoting studies based on simulation results and ferromagnetic chains, respectively. The second column describes the uniaxial magnetic anisotropy term $D$, either considered equal for all atoms or uniquely defined for each atom (three cases of trimers). The third column describes the transverse magnetic anisotropy parameter $E$ and has one study with a unique definition for each individual atom of a trimer. Ref.~\cite{RudowiczApplMagnRes2019}
    found this to be negative, contrary to our definition. The fourth column describes the Heisenberg exchange coupling strength $J$, showing both ferromagnetic ($J<0$) and antiferromagnetic ($J>0$) couplings. A study~\cite{NicklasJApplPhys2011} reporting $J$ larger than $25$~meV is inconsistent with our observations. Ref.~\cite{Rolf-PissarczykPRL2017} describes different coupling strengths between different neighbours in a trimer. The fifth column describes the g-factor. The sixth and seventh columns provide the reference and publication year of these studies, respectively.}
    \label{tab:OverviewOfParameters}
\end{table}

Table \ref{tab:SimulationParameters} provides a summary of the parameters used in the simulations of each figure. The simulation result used for Fig.~1a was processed by subtracting a linear slope along $B_x$ for clear visualization of the diabolic point. In the other simulations, we considered angles, $\alpha$ and $\beta$, representing the directions of the applied magnetic field with respect to the crystal axes (further details in Supplementary Note 3). In addition to the external magnetic field, the tip magnetic field induced by the magnetic interaction between the magnetic tip and the atom beneath it was considered. Note that we considered the $B_z$ component only, given the negligibly small tip field along $B_x$ compared to the external field. The coupling with the substrate is given as $G_{\rm SS} = 1$~$\mu$S, following the definition given in \cite{Rolf-PissarczykPRL2017}. \\

Noting that the simulation shows lifetimes larger than the experimental values, especially closer to the peak center, we believe the measured lifetime is limited by accidental voltage spikes that may cause over-the-barrier transitions. This is corroborated by longer lifetimes measured upon better grounding of thermometry lines close to the current line. While we did not find a proper way to include such spurious voltages in the model, the simulations were performed with elevated temperatures to account for the on-average higher electron energy. We believe there has been a small voltage offset during the measurements of Fig.~3a, resulting in voltages slightly below the intended values. For Fig.~2 and 3b, we introduced a relatively strong tip polarization $\eta$ (defined between $-1$ and $+1$), which gives better match between the simulation and experimental results. To expedite the processing time, only the lowest 250 energy eigenstates were considered, with only the 500 largest contributions of spin eigenstates within these energy eigenstates. \\

\begin{table}
    \resizebox{\textwidth}{!}{
    \begin{tabular}{c|ccccc}
        Paramater & Figure 1a & Figure 2d & Figure 2e & Figure 3a & Figure 3b \\
        \hline
        $N$                  & 1     & 5                                       & 5 & 6 & 5  \\
        $D$ (meV)            & -1.87 & $\{-2.05, -2.35, -2.85, ...\}$ & $\{-2.1, -2.3, -2.85, ...\}$ & $\{-1.95, -2.25, -2.80, ...\}$  & $-1.70$ (all) \\
        $E$ (meV)            & 0.31  & $\{0.33, 0.32, 0.32, ...\}$      & $0.32$ (all) & $ 0.3$ (all) & $0.4$ (all) \\
        $J$ (meV)            &       & $\{0.85, 1, ...\}$                  & $ 0.83$ (all) & $ \{0.85, 0.95, 1.35... \}$ & $-0.95$ (all) \\
        $g$                  & 2.11  & $\{2.5, 2.11, 2.11, 2.11, 2.11\}$       & $\{2.5, 1.9, 2.5, 1.9, 1.8\}$ & $\{2.4, 2.11, 2.11, 2.11, 2.11, 2.11\}$ & $ \{2.5, 2.11, 2.11, 2.11, 2.11\}$  \\
        $B_{\rm tip}$ (mT) &       & 50 on atom i                            & 100 on atom i & -110 on probed atom & -60 on atom i \\
        $\alpha (^{\circ})$ &       & 0.2                                     & 0.8 & 4.1 & 0.2 \\
        $T$ (K)              &       & 3                                       & 1.3 & 2.3 & 1.3 \\
        $V$ (mV)             &       & 3                                       & 3 & 2.75 (purple), 4.75 (grey) & 3 \\
        $G_{\rm SS}$~($\mu$S)     &       & 1                                       & 4   & 4 & 4 \\
        $\eta$             &       & 0.5                                       & 0.5   & 0 & 0.5 \\
    \end{tabular}}
    \caption{\textbf{Overview of parameters used in the simulations presented in the main text.} $D$, $E$, $J$, and $g$ are the same parameters as given in Table~1. $B_{\rm tip}$ represents the magnetic field induced on a single atom due to the tip field along $B_z$, $\alpha$ the angle between the external magnetic field and the crystal lattice, $V$ the voltage applied for measurements, $G_{\rm SS}$ the electron coupling with the bath, and $\eta$ the polarization of the tip. For $D$, $E$, $J$, and $g$, a list of values is given, corresponding to the values on or between the various atoms, starting with atom I. Here an ellipsis ("...") indicates that the remaining values are symmetric with respect to the previous values (following $a,b,c,b,a$ for $N = 5$, and $a,b,c,c,b,a$ for $N = 6$).}
    \label{tab:SimulationParameters}
\end{table}

\section*{Supplementary Note 2: Simulations of lifetime around a diabolic point}
\vspace{5mm}
\addcontentsline{toc}{section}{Supplementary Note 2: Simulations of lifetime around a diabolic point}

The simulations to describe the evolution of lifetime as a function of the external magnetic fields are based on rate equations~\cite{Rossier_prl_2009,Delgado_Palacios_prl_2010,LothNaturePhysics2010}, which include transitions between various states following spin-selection and energy conservation rules. These transitions can be derived by including the exchange coupling of the local spins with the itinerant electrons \cite{DelgadoProgSS2017}. In a semiclassical description, under an applied longitudinal field $B_z$, a spin $S$ with an easy-axis anisotropy exhibits the ground state $\psi_0$ with $m_s=-S$ along the easy axis and the first excited state $\psi_1$ with $m_s=+S$. Thus transitions between these states while exchanging spin angular momentum with tunneling electrons necessitate passing through intermediate states, adhering to the spin conservation rule ($\Delta m_S=\pm 1$). These intermediate states are higher in energy, which requires tunneling electrons with sufficient energy to overcome the effective energy barrier. 
In a quantum mechanical framework, states $\psi_0$ and $\psi_1$ experience slight hybridization, leading to direct transitions between them without exchanging the spin angular momentum ($\Delta m_s = 0$). For a single Fe atom on Cu$_2$N, the transverse magnetic anisotropy induces this hybridization, resulting in the ground state comprising a linear combination of spin states $m_s = -2$ and $m_s = +2$. This hybridization facilitates Quantum Tunneling of Magnetization (QTM), leading to direct transitions between both states. \\

We can thus differentiate between two types of transitions. First, there is the so-called flip-flop transition where the magnetic angular momentum of the tunneling electron is exchanged with the one of the atom, leading to $\Delta m_s = \pm 1$, which is the primary cause of over-the-barrier transitions and is expressed as $\langle\psi_0|S_{\pm,i}|\psi_1\rangle$. Secondly, there is a so-called direct transition, where $\Delta m_s = 0$, which is the primary cause of QTM. By suppressing the hybridization, the direct transitions, whose intensities depend on $\langle\psi_0|S_{z,i}|\psi_1\rangle$, are suppressed. This quenching reduces the switching rates due to QTM, consequently leading to longer lifetimes in the absence of over-the-barrier excitations. The transition likelihood between $\psi_0$ and $\psi_1$ for a single tunneling electron can be encapsulated by the scattering intensity $P_{0,1} =   \sum_{a=x,y,z}|\langle \psi_0|S_{a,i}|\psi_1\rangle|^2$, which is plotted in Figs.~2d, 2e, 3a and 3b and is a measure for how likely QTM is. \\

\begin{figure}[H]
\centering
\includegraphics[width=1\linewidth]{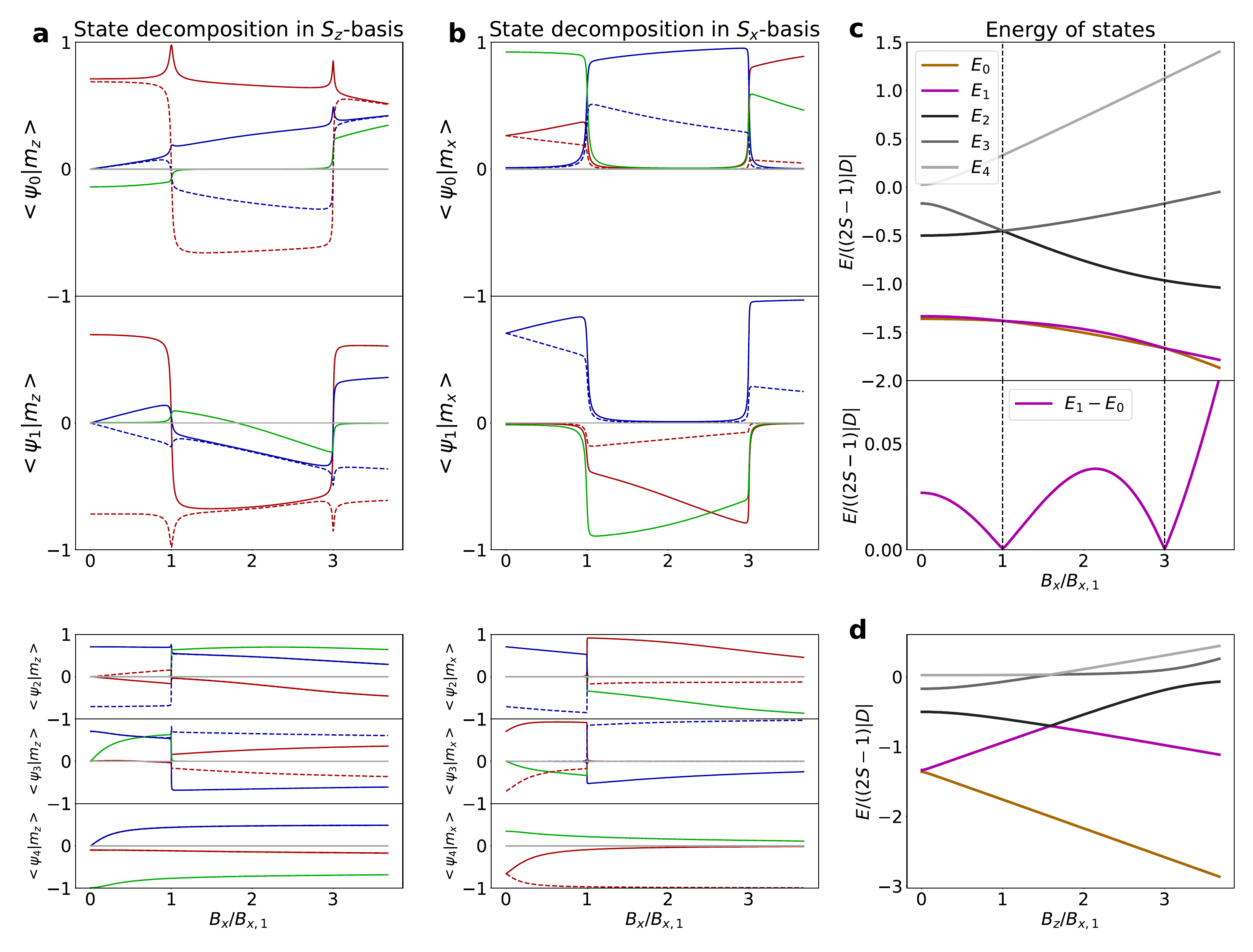}\caption{\textbf{State decomposition of the single Fe atom as a function of $B_x$.} \textbf{(a)} Decomposition of $\psi_0$ and $\psi_1$ (top) in the $S_z$-basis, and $\psi_2$ through $\psi_4$ (bottom) as a function of $B_x$ divided by $B_{x,1}$, the field value of the first diabolic point. \textbf{(b)} Same but in the $S_x$-basis. Red curves represent $m_s = \pm 2$, blue curves correspond to $m_s = \pm 1$, and a green curve is for $m_s = 0$ in the respective basis. Dashed lines are for $m_s > 0$. \textbf{(c)} (top) Energies of the five states of the single Fe atom over the same magnetic field range as panels a and b. The abrupt changes in the decomposition curves around $B_{x,1}$ and $3B_{x,1}$ correspond to avoided level crossings in the energy spectrum. (bottom) Energy difference between $\psi_0$ and $\psi_1$ from the top panel. \textbf{(d)} Energies of the five states when the magnetic field is applied along $B_z$ instead. In all cases, there is $10$~mT in the other in-plane direction to induce smooth avoided level crossings. }\label{fig:S:Decomposition}
\end{figure}

More specifically, in the absence of an external magnetic field, hybridization occurs from the transverse magnetic anisotropy term $E(S_x^2 - S_y^2)=E(S_+^2 + S_-^2)/2$. For a single Fe atom with $S=2$, the magnetic anisotropy terms, $D = -1.87$~meV and $E = 0.31$~meV, result in the ground state of $\psi_0 = (0.7, 0, -0.139, 0, 0.7)$ as a symmetric superposition of $m_s = \pm 2$. The first excited state $\psi_1 = (0.707, 0, 0, 0, -0.707)$ represents an antisymmetric superposition. Figure~\ref{fig:S:Decomposition}a and b show how these states evolve when sweeping $B_x$ as given by the $S_z$ and $S_x$ basis, respectively. \\

Around $B_{x,1}$ and $3B_{x,1} = B_{x,3}$, there are avoided level crossings between $\psi_0$ and $\psi_1$ (Fig.~\ref{fig:S:Decomposition}c), where the two states effectively swap their spin composition. In terms of $S_z$-basis, this results in a switching of the symmetry, i.e., for the first avoided level crossing, a transition from $|\psi_0\rangle\sim (|-2\rangle_z+|+2\rangle_z)/\sqrt{2}$ to $|\psi_0\rangle\sim (|-2\rangle_z-|+2\rangle_z)/\sqrt{2}$.
In the process, the $\langle\psi_0|+2\rangle_z$ contribution goes through zero while the  $\langle\psi_0|-2\rangle_z$ one increases in contribution. A similar effect happens at the second avoided level crossing. In both cases, at the exact crossing points, $\psi_0$ has zero contribution of $m_z = +2$ and $\psi_1$ has zero contribution of $m_z = -2$. This significantly suppresses the $\Delta m_z = 0$ transition probability between the two states, compared to field values far from these crossings. The width of the lifetime peak vs. $B_x$, or, alternatively, the width of the avoided level crossing in the energy spectrum (see Fig.~\ref{fig:S:Decomposition}c) is ultimately determined by the $B_z$-field, being strictly zero when $B_z=0$. For a finite $B_z$, the hybridization between the $|\pm 2\rangle_z$ states is reduced, and the quenching of the QTM near a DP survives in a larger $B_x$-window. This quenching also happens between $\psi_2$ and $\psi_3$, although the effect is much sharper as the $B_x$ field has a much more drastic effect on the energy splitting between the $\psi_2$ and $\psi_3$ states. \\

While the $S_z$-basis is convenient for explaining the critical behaviour of lifetimes near the diabolic point, a better insight into the underlying mechanism can be gained using the basis of spin states along the applied field, i.e., the $S_x$-basis (Fig.~\ref{fig:S:Decomposition}b). For small $B_x$, $\psi_0$ is dominated by $m_x = 0$ with minor contribution in $m_x = \pm 2$ due to the transverse magnetic anisotropy. At the first diabolic point, $\psi_0$ and $\psi_1$ swap their spin compositions. Thus, $\psi_0$ converts from a state dominated by $m_x = 0$ to one dominated by $m_x = -1$. Then, at the second diabolic point, $\psi_0$ converts into a state dominated by $m_x = -2$. Hence, at each diabolic point, a quantum of $S_x$ enters the ground state. As shown in Fig.~\ref{fig:S:Decomposition}, abrupt changes in both energy spectra and states' composition only appears near the DPs. \\

In summary, we can understand each crossing between $\psi_0$ and $\psi_1$ as the change of a quantum of $S_x$ in the ground state. Thus, as $B_x$ is swept, the ground state changes from the majority state of $m_x = 0$ to $m_x = -1$ after the first crossing and to $m_x = -2$ after the second crossing. In terms of the $S_z$ state composition, an additional unit of $S_x$ results in alternating symmetry in the linear combinations of $m_z = \pm m$ for any $m$. This swapping of symmetry results in values of $B_x$ where the hybridization is minimized, ultimately leading to maxima in the lifetimes. \\

\begin{figure}[H]
\centering
\includegraphics[width=1\linewidth]{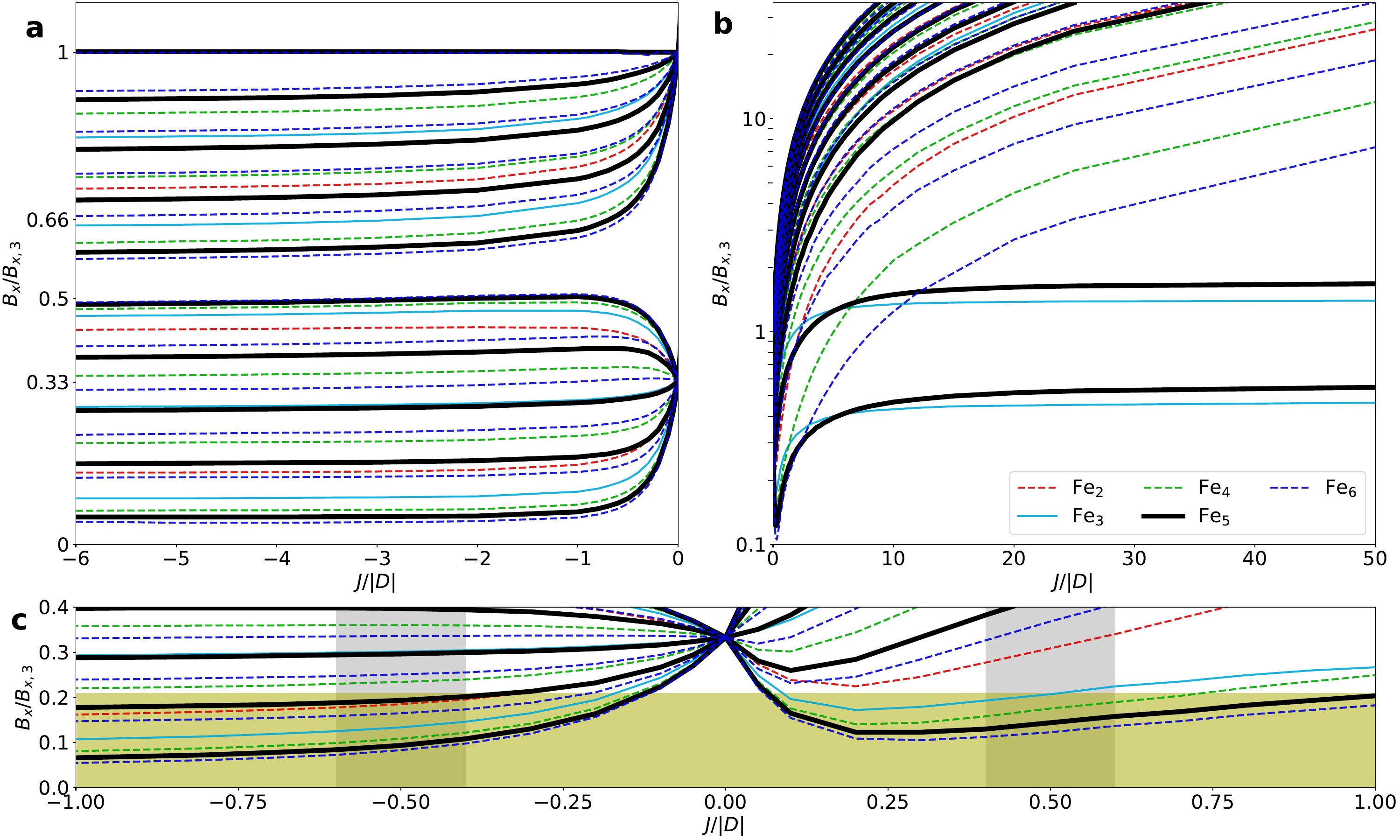} 
\caption{\textbf{Magnetic field values of diabolic points for Fe chains with different lengths $N = \{2,3,4,5,6 \}$ and different coupling strengths $J/|D|$.} \textbf{(a)} Ferromagnetically coupled atomic chains ($J/|D| \le 0$) in a linear scale. \textbf{(b)} Antiferromagnetically coupled case  ($J/|D| > 0$) in a logarithmic scale to facilitate comparison under different regimes. \textbf{(c)} Zoom-in for $|J/D| \le 1$ and $B_x \le 0.4B_{x,3}$. The shaded region contains values of $B_x$ and $J/|D|$ relevant to the experiments performed throughout this work. The horizontal shaded region indicates our external magnetic field range. All plots were constructed by finding degeneracies between $\psi_0$ and $\psi_1$ as a function of $B_x$ (here, $B_z=0$). The legend presented in panel b applies to all panels. To facilitate comparison, solid lines (dashed lines) indicate odd (even) number chains. }\label{fig:S:JN}
\end{figure}

Compared to the single Fe atom, the behaviour of spin chains is more intricate but fundamentally based on the same physics for weak coupling between the atoms. The number of diabolic points grows with the total spin of the system and, consequently, with the number of atoms $N$. This is illustrated in Fig.~\ref{fig:S:JN} where, using the same values of $D$, $E$, and $g$, we calculated all positive diabolical points $B^{N,j}_{x}$ for the $j \in \{1,...,NS\}$ values of chains with $N=2-6$ spins for different coupling strengths of $J$ including both ferromagnetic ($J<0$, Fig.~\ref{fig:S:JN}a) and antiferromagnetic ($J>0$, Fig.~\ref{fig:S:JN}b) cases. The $N = 5$ case is shown in thick black lines for clarity. The value $j = 1$ corresponds to the lowest DP. In Fig.~\ref{fig:S:JN}c, we narrow our focus to a smaller window aligned with experimental conditions.

For $J = 0$, the $N$ atoms are decoupled, resulting in the same diabolic points at $B_{x,1}$ and $B_{x,3}$. As $|J|$ increases, the degeneracy between atoms is lifted, leading to the dispersion of diabolic points into multiple branches. The number of these branches corresponds to the number of atoms in the chain. Consequently, by increasing $j$, the quanta of $S_x$ in the ground state increase upon passing the $j$\textsuperscript{th}-DP, as happened twice for Fig.~3b. \\

There is a major difference between the ferromagnetic and antiferromagnetic chains. The ferromagnetic case (Fig.~\ref{fig:S:JN}a) is certainly more intuitive: as $-J/|D|$ increases, the diabolic points tend towards the DPs of a macrospin of dimension $NS$. This can be observed in the spacing between each subsequent diabolic point, which follows the same spacing as described in Eq.~2 of the main text. Surprisingly, irrespective of the length of the chain, the highest diabolic point is always equal to $B_{x,3}$. This implies that the lowest diabolic point can be found at $B_{x,3} / (2SN - 1)$, and the next diabolic point can be found at three times this value. The ferromagnetic chain in the main text has a ratio $B^{5,2}_{x}/B^{5,1}_{x}\sim 2$, suggesting that $|J|$ is too small to consider the chain as a single macrospin, consistent with our simulations. \\

For the antiferromagnetic ($J > 0$) chain (Fig.~\ref{fig:S:JN}b), when $J/|D| \gg 1$, the odd-numbered chains present $S$ DPs at $B_x\lesssim B_{x,2S-1}$. This corresponds to the uncompensated spin in such chains. This may also open the possibility of exploring diabolic points in Mn$_N$-chains as the ones explored by Hirjibehedin {\em et al.} \cite{HirjibehedinScience2006}, where $J/|D|\approx 160$ \cite{HirjibehedinScience2007} and $B_{x,1}\approx 0.5$ T. Simulations show lifetime peaks at around $0.75$~T and $1.5$~T for Mn$_3$ and at around $1$~T and $2$~T for Mn$_5$. \\

In agreement with our experimental observations, Fig.~\ref{fig:S:JN}c shows that as $N$ increases for values $|J|/|D| \approx 0.5$, the lowest diabolic point $B^{N,1}_{x,1}$ is found at smaller values of $B_x$. For much larger $J$, all DPs of even-numbered antiferromangetic chains diverge to extreme values. It also shows that in our operable window (up to about $0.21 B_{x,3}$), the easiest way to measure multiple DPs is by going to ferromagnetic chains. We had considered antiferromagnetic chains of longer length, like Fe$_8$. However, the time needed to measure statistically relevant data at the DP exceeded the scope of this work. \\

The only noteworthy effect of local changes in the $g$-factors is a small splitting of the $B^{N,j}_{x}$ diabolic points when $|J/D| \rightarrow 0$, not shown in the figure. 

\newpage

\section*{Supplementary Note 3: Sample orientation}
\vspace{5mm}
\addcontentsline{toc}{section}{Supplementary Note 3: Sample and magnetic field orientation}

\begin{figure}[H]
\centering
\includegraphics[width=0.9\linewidth]{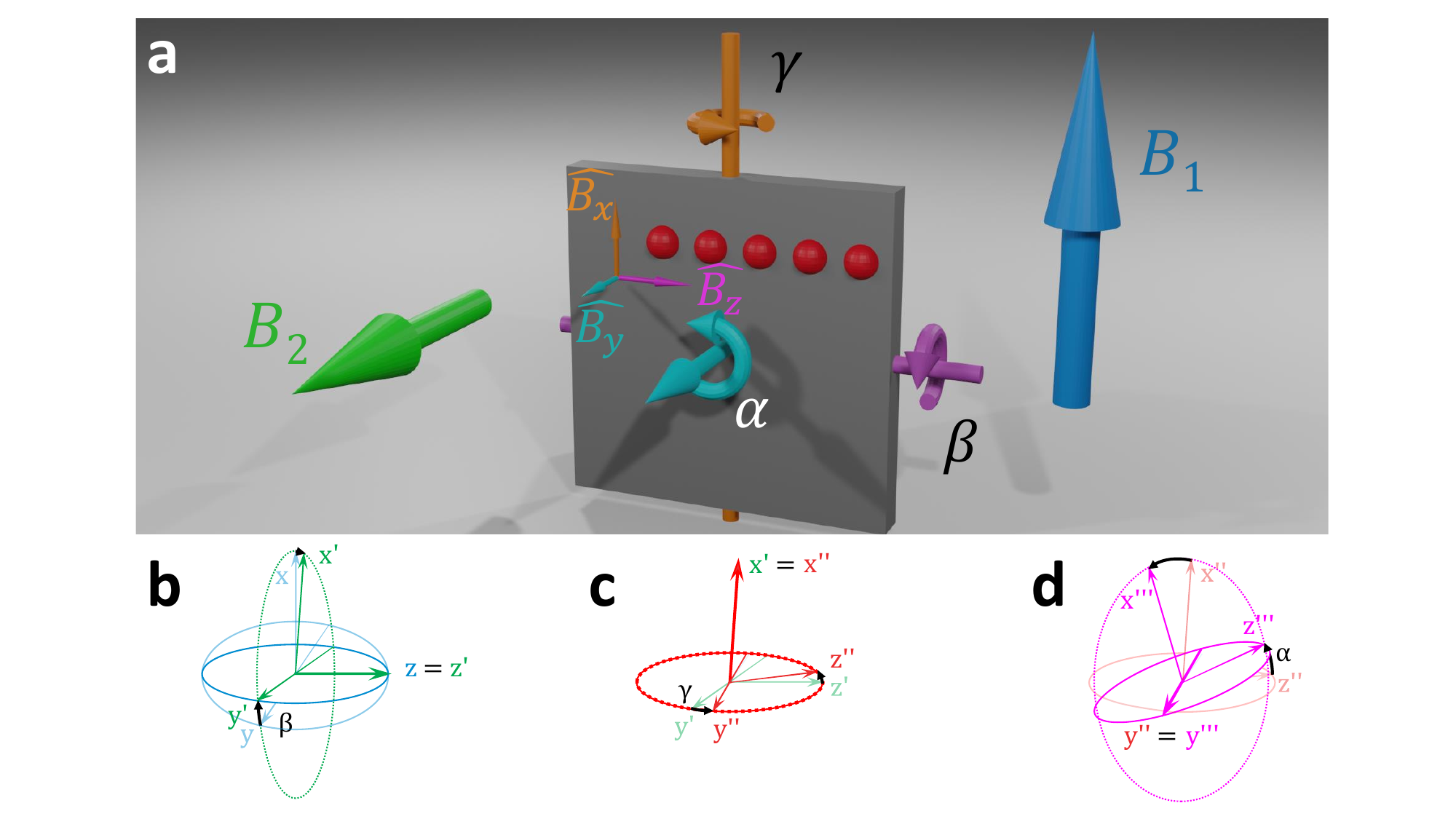}\caption{\textbf{Alignments of the sample with magnetic fields.} \textbf{(a)} The substrate is given as a square plate, on which five red spheres represent the Fe$_5$ atom chain. The substrate's orientation can be adjusted along three axes through angles $\alpha$, $\beta$, and $\gamma$. An external magnetic field, facilitated by a two-axis vector magnet, can be applied along $\boldsymbol{B}_1$ and $\boldsymbol{B}_2$. Due to misalignment between the crystal axes and the magnetic field directions, the applied fields can be decomposed into $\boldsymbol{B}_x$, $\boldsymbol{B}_y$ and $\boldsymbol{B}_z$ for the chains in terms of crystal axes. \textbf{(b)} Tait-Bryan rotation along the z-axis with an angle $\beta$. A negative value of $\beta$ is shown to represent the actual angle $\sim -8^\circ$. \textbf{(c)} Tait-Bryan rotation along the x'-axis with an angle $\gamma$. A small value $\gamma$ is shown to represent the near negligible actual value. \textbf{(d)} Tait-Bryan rotation along the y''-axis with an angle $\alpha$. A large value of $\alpha$ is shown to represent the actual large variability of $\alpha$ in the work presented here. }
\label{fig:S:SampleOrientation}
\end{figure}

Since the shape of the lifetime curves strongly depends on the exact orientation and magnitude of the applied magnetic field, particularly the $B_z$ component, the precise characterization of magnetic field components with respect to the crystal axes is crucial. Figure \ref{fig:S:SampleOrientation} shows the Fe$_5$ chain on a Cu(100) crystal, illustrating the directions of the external magnetic field and the crystal axes. In our experiment, we found that the crystal axes are not fully aligned with the external magnetic field directions due to several reasons: \\

\begin{enumerate}
    \item The sample holder, designed for a certain angle ($\sim 8^\circ$) for the cold deposition \cite{HwangRSI2022}, contributes significantly to the misalignment between $\boldsymbol{B}_1$ and $\boldsymbol{B}_x$, resulting in the angle $\beta = \sim -8^\circ$.
    \item Mounting the Cu(100) crystal to the sample holder introduces a slight in-plane rotation, which gives a finite $\alpha$. This can be estimated from the atomic resolution images and adjusted by remounting the crystal (Fig.~\ref{fig:S:Tilt}b,d,f). 
    \item There is a subtle misalignment between the STM stage and the magnetic field axes, which affects all three angles. From the experiment, we found the effects on $\gamma$ and $\beta$ are negligibly small while giving $ \sim 3^\circ$ deviation for $\alpha$, denoted as $\alpha_{\rm tilt}$.
\end{enumerate}

The resulting combination of rotations can be expressed in a sequence of Tait-Bryan rotations, as indicated in Fig.~\ref{fig:S:SampleOrientation}b-d. Applying the rotations in the given order results in the following decomposition of the external field:

\begin{equation}
\begin{aligned}
    B_x = & B_1(\cos \alpha \cos \beta - \sin \alpha \sin \beta \sin \gamma) &+ B_2 (\cos \alpha \sin \beta + \sin \alpha \cos \beta \sin \gamma) \\
    B_y = & B_1(- \sin \beta  \cos \gamma) & + B_2 (\cos \beta \cos \gamma) \\
    B_z = & B_1(\sin \alpha \cos \beta + \cos \alpha \sin \beta \sin \gamma) &+ B_2 (\sin \alpha \sin \beta - \cos \alpha \cos \beta \sin \gamma) \\
\end{aligned}
\end{equation}

\noindent  In the limit where $\gamma = 0$, $B_2 = 0$, $|\beta| \ll \pi/2$ and $|\alpha| \ll \pi/2$, these equations can be simplified to: 

\begin{equation}
\begin{aligned}
    B_x = & B_1(\cos \alpha \cos \beta)  \\
    B_y = & B_1(- \sin \beta ) \\
    B_z = & B_1(\sin \alpha \cos \beta) \\
\end{aligned}
\end{equation}

\section*{Supplementary Note 4: Variation of lifetime curves at different conditions}
\vspace{5mm}
\addcontentsline{toc}{section}{Supplementary Note 4: Variation of lifetime curves at different conditions}

\begin{figure}[H]
\centering
\includegraphics[width=1\linewidth]{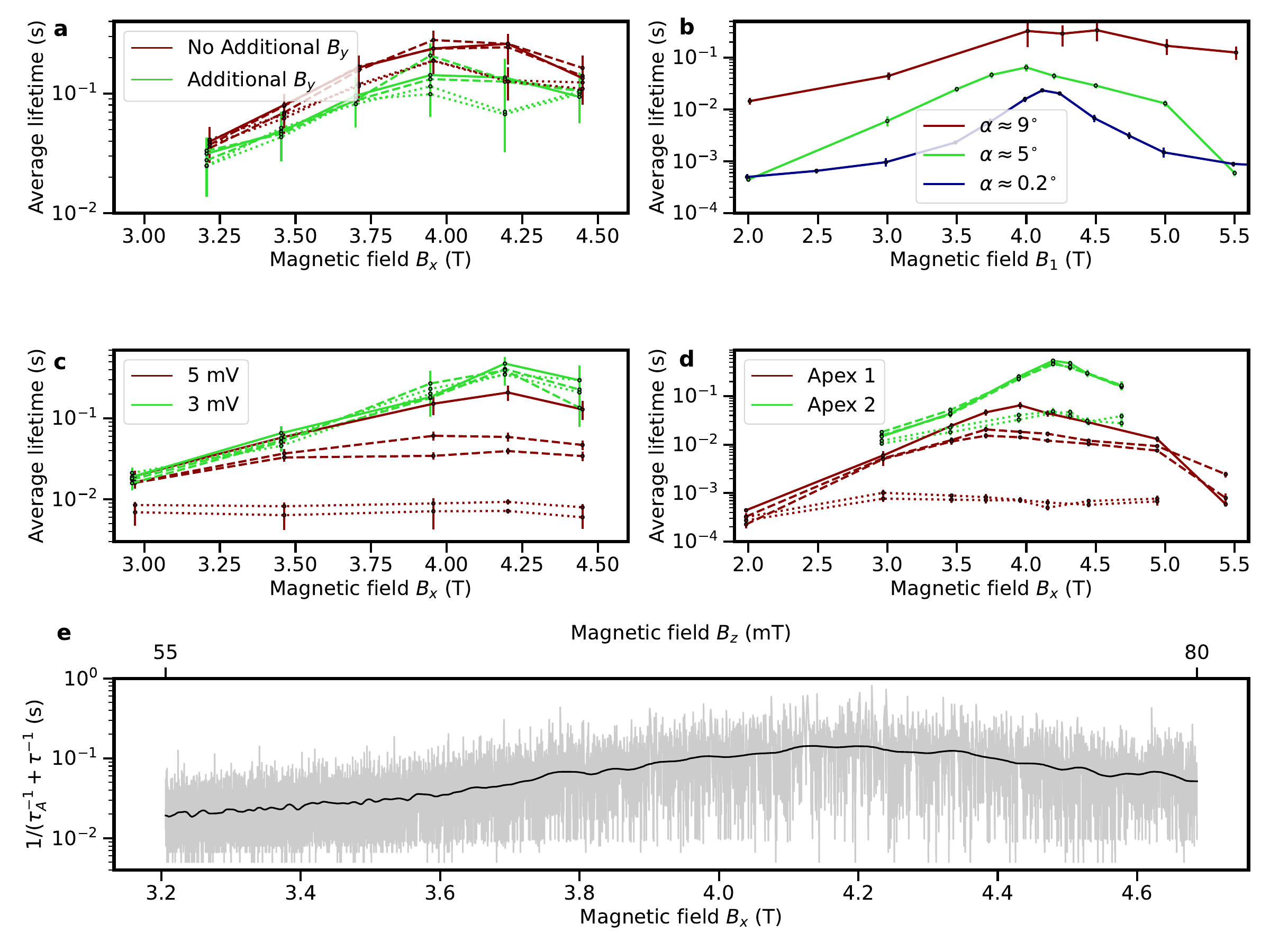}
\caption{\textbf{Lifetime curves for antiferromagnetic Fe$_5$ under various experimental conditions.} \textbf{(a)} Lifetimes measured for each Fe atom in the chain at different out-of-plane fields, $\boldsymbol{B}_y$. The variation in $\boldsymbol{B}_y$ is primarily achieved by adjusting the $\boldsymbol{B}_2$ field. For the green curves, $\boldsymbol{B}_2 = \sin(8^{\circ})B_1$, while for the red curves, $\boldsymbol{B}_2 = 0$. \textbf{(b)} Lifetimes measured for atom III at different $\alpha$ values obtained by rotating the crystal. The labels indicate the actual angle. The data in the main text (Fig~2d) is included in blue. \textbf{(c)} Lifetimes for each Fe atom in a chain measured at different bias voltages. \textbf{(d)} Lifetimes measured under the same conditions on the same chain but using two different tips. \textbf{(e)} Up-to-up switching time, defined as $1/(1/\tau_A + 1/\tau_B)$ (gray), while sweeping the external magnetic field from $4.75$~T to $3.25$~T. During the sweep, the tip remained in constant-current mode over the atom III to compensate for drift. The time between switches was long enough such that the tip height followed telegraphic noise-like behaviour. A rolling average of 50 points is given in black. All magnetic field values, $B_x$ and $B_z$, were determined by converting the applied external magnetic fields considering the crystal axes. All data (except for the blue curve in (b) and the green curve in (c)) were obtained with $V_{\rm bias} = 5$~mV, $I = 10$~pA, $T = 1.3$~K (blue data obtained at $T = 1.4$~K and $V = 3$~mV). All data (except the red and blue curves in (b)) were obtained with $\alpha = 5^{\circ}$. Error bars represent 2 standard deviations. Connecting lines between data points are guides to the eye. Solid lines correspond to atom III, dashed lines to atoms II and IV, and dotted lines to atoms I and V.}\label{fig:S:Robust}
\end{figure}

To clarify major effects of $B_y$ (Fig.~\ref{fig:S:Robust}a), $\alpha$ (Fig.~\ref{fig:S:Robust}b), $V_{\mathrm{bias}}$ (Fig.~\ref{fig:S:Robust}c) and tip apex (Fig.~\ref{fig:S:Robust}d) on the shape of the lifetime curves, we present further data sets obtained under various conditions on an antiferromagnetic Fe$_5$ chain. Further analyses of voltage dependence, as well as current dependence, are explored in Supplementary Notes 7 and 8, respectively.  \\

As explained in the previous section, although the external magnetic fields $\boldsymbol{B}_1$ and $\boldsymbol{B}_2$ are closely aligned along the in-plane and out-of-plane directions, respectively, simultaneous adjustments of $\boldsymbol{B}_1$ and $\boldsymbol{B}_2$ are necessary to selectively apply a field along one of $\hat{B_x}$, $\hat{B_y}$ and $\hat{B_z}$. For instance, applying $\boldsymbol{B}_1$ to change the in-plane field induces a finite out-of-plane field, given by $B_y = B_1 \sin(8^\circ)$. To understand the effect of the out-of-plane magnetic field on the lifetime, we varied $\boldsymbol{B}_2$ in a way to double the total $B_y$ experienced by the chain, i.e., $B_2 = B_1 \sin(8^\circ)$. As shown in Fig.~\ref{fig:S:Robust}a, the overall lifetimes show a slight reduction, and the peak near the diabolic point is slightly attenuated with increased $B_y$, consistent with simulations. Although a more thorough analysis is required, we expect that fully compensating $B_y$ could yield slightly more pronounced peaks near the diabolic point. However, since the influence of $B_y$ is not substantial and a small $B_y$ would not change the underlying physics behind our studies, we kept $B_2=0$ for the rest of our experiments. \\

Adjusting the angle $\alpha$ is correlated with changing the ratio between two in-plane fields, $B_x$ and $B_z$. A larger $\alpha$ corresponds to a higher $B_z$ at a certain $B_1$. The increase in $B_z$ is associated with an increasing contribution of the main N\'eel states for $\psi_0$ and $\psi_1$, at the expense of the minor contribution. In other words, $\psi_0 \rightarrow N_A$ and $\psi_1 \rightarrow N_B$, leading to a reduction in hybridization. Thus, at a larger $\alpha$, we observed an overall increase in the lifetime compared to a smaller $\alpha$. Secondly, the Zeeman splitting between these two lowest-lying states was increased, which results in the smearing out of the peak near the diabolic point (Fig.~\ref{fig:S:Robust}b). Note that the lifetime curves were measured with different samples and different tips since we needed to remount the crystal to rotate it and thus to change $\alpha$, which may introduce additional effects of local environments on the chain, such as local defects, strain in the underlying Cu$_2$N layer, diverse tip apexes, and so on. The influence of different tip apexes is shown in Fig.~\ref{fig:S:Robust}d. As measured on the same Fe$_5$ chain but with different spin-polarized tips, the lifetime varies not only in the magnitudes but also in the magnetic field value for the lifetime peak. This indicates variations of the tip magnetic fields along both $B_x$ and $B_z$. \\

Figure \ref{fig:S:Robust}c shows the effect of the applied bias voltages. Increasing the voltage results in more over-the-barrier transitions and, thus, larger variations in the lifetime among the atoms in the chain (consistent with the data shown in Fig.~3a of the main text). See also Supplementary Notes 8 and 9 for further discussions. \\

Figure \ref{fig:S:Robust}e shows the up-to-up switching times during a magnetic field sweep. The raw data is given in gray, while the black line shows a rolling average using SciPy's 1D Gaussian filter (50 data points). Even for this rough measurement, we can clearly see the switching rate is significantly reduced near 4.15~T, corresponding to the diabolic point. Considering that obtaining a dataset for lifetime curves takes at least one day, this 45-minute measurement provides a quick way to verify the location of the diabolic point. \\

\section*{Supplementary Note 5: Symmetry breaker of Fe$_6$}
\vspace{5mm}
\addcontentsline{toc}{section}{Supplementary Note 5: Symmetry breaker of Fe$_6$}

In the absence of external magnetic fields, the two lowest-lying eigenstates, $\psi_0$ and $\psi_1$, in an Fe chain are composed of the symmetric and antisymmetric superposition of the N\'eel states, respectively. The energy difference between these states arises from transverse magnetic anisotropy ($E(S_x^2 - S_y^2)$). This energy difference is effectively mitigated by applying a transverse magnetic field $B_x$. At the diabolic point, the $B_x$ fully compensates for the energy difference, leading to an energy level crossing. In an ideal situation, where all atoms on a surface are identical and devoid of any variations in local environments, the diabolic point is expected to manifest in an exceedingly sharp window of $B_x$ (a singular point value) with nearly infinite lifetimes. This sharp transition can be smeared out by converting the level crossing into an avoided-level crossing, where the broadening is proportional to the energy difference between the two states. \\

For a ferromagnetic chain or an odd-numbered antiferromagnetic chain, this avoided-level crossing can be easily induced by applying a longitudinal magnetic field $B_z$, which yields the energy difference between two N\'eel states by the Zeeman energy. In contrast, achieving an avoided-level crossing for an even-numbered antiferromagnetic chain is nontrivial, as the two N\'eel states possess identical energies across all $B_z$ fields. This conflicts with our observations from the Fe$_6$ chain in Fig.~3a, which suggests a symmetry break between the two N\'eel states, ultimately making one state more favored than the other under a finite Zeeman energy. We attribute this symmetry break to inhomogeneity in the chain. In reality, there are subtle variations in local environments among atoms on surfaces, which results in different $g$-values and magnetic interactions between them. In addition, the presence of a magnetic tip located over one atom in the chain provides a tip-induced local magnetic field. \\

In our simulations, the Zeeman energy induced by variations in $g$-values is included by assigning a different $g$-value to atom I in the chain rather than introducing varied $g$-values for all atoms. Consequently, the total Zeeman splitting is proportional to an applied field along the $\hat z$-axis and the $g$-factor mismatch between the atom I and the rest of the atoms $g_1 - g_{\mathrm{rest}}$. \\

\begin{figure}[H]
\centering
\includegraphics[width=0.9\linewidth]{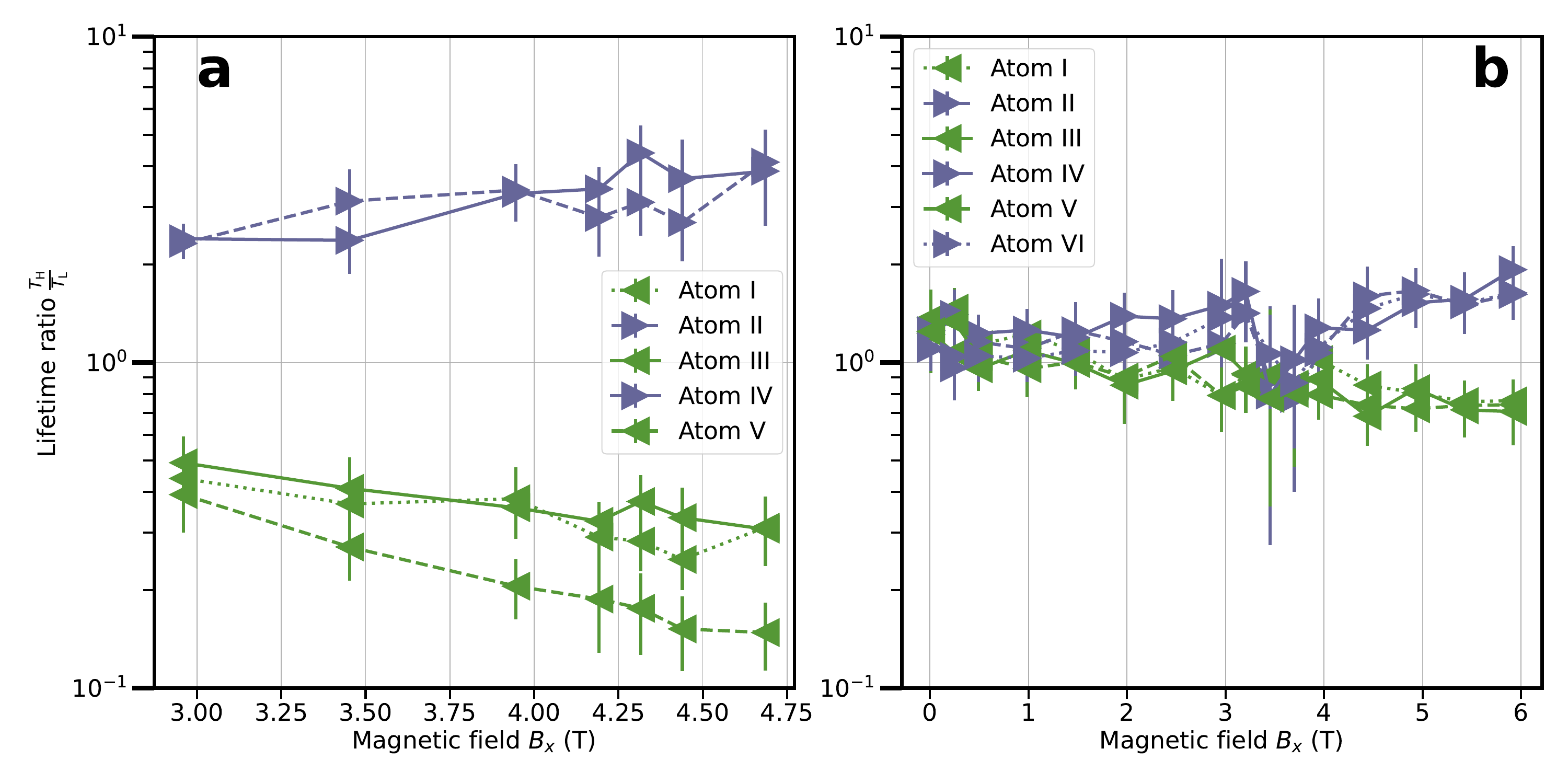}
\caption{\textbf{Lifetime ratios defined as $T_{\mathrm H}/T_{\mathrm L}$ for odd- and even-numbered chains}.
\textbf{(a)} Lifetime ratios of a Fe$_5$ chain obtained at $\alpha = 5^{\circ}$. In the chain, even-numbered atoms II and IV exhibit a lifetime ratio of about 3, while odd-numbered atoms I, III, and V show a lifetime ratio of about $1/3$ across the given magnetic field range. \textbf{(b)} Lifetime ratios of Fe$_6$ for the data presented in the main text. Blue and green colors indicate even- and odd-numbered atoms, respectively. A split between even- and odd-numbered atoms starts to emerge around $2$~T, diminishes around the diabolic point, and increases for higher magnetic fields. Lines are guides to the eye. Error bars represent two standard deviations. Data in panel a corresponds to the $10$~pA data presented in Fig.~\ref{fig:S:Current}. }\label{fig:S:SymBreaker}
\end{figure}

To confirm the symmetry break, we compare the lifetime ratios between atoms in the Fe$_5$ and Fe$_6$ chains. While our primary focus in the main text is on $T_{\mathrm A}$ and $T_{\mathrm B}$ for the average lifetimes associated mostly with N\'eel states $N_{\mathrm A}$ and $N_{\mathrm B}$, respectively, we shift our attention here to the average lifetimes of $T_{\mathrm H}$ and $T_{\mathrm L}$ related to high and low spin-polarized current states, respectively. This keeps the analysis more general and highlights the alternating pattern of antiferromagnetic chains better. At the diabolic point, in the absence of Zeeman energy, the two lowest-lying states are degenerate, resulting in equal lifetimes for $T_{\mathrm H}$ and $T_{\mathrm L}$ (i.e. $T_{\mathrm H}/T_{\mathrm L} \approx 1$). Importantly, this means that a larger imbalance between $T_{\mathrm H}$ and $T_{\mathrm L}$ indicates a greater energy difference between the two states. Thus, we use the lifetime ratio $T_{\mathrm H}/T_{\mathrm L}$ as a proxy for the ratio of state ensemble occupation, which, according to the Boltzmann distribution, provides insights into the energy difference between the two states.
\\

Figures~\ref{fig:S:SymBreaker}a and b present the lifetime ratios ($T_{\mathrm H}/T_{\mathrm L}$) for the Fe$_5$ and Fe$_6$ chains, respectively. In the Fe$_5$ chain, the lifetime ratio for atoms I, III and V is about $1/3$, while for atoms II and IV, it is about $3$. This simple inversion in the $T_{\mathrm H}$ and $T_{\mathrm L}$ ratios indicates that $\psi_0$ and $\psi_1$ predominantly consist of specific N\'eel states, namely $N_{\mathrm A}$ and $N_{\mathrm B}$, respectively. Interpreted as population ratios, the lifetime ratios given in Fig.~\ref{fig:S:SymBreaker}a correspond to an energy difference between $\psi_0$ and $\psi_1$ of about $140$~$\mu$eV at around 4T. Note that this dataset was obtained with the sample's angle of $\alpha \approx 5^{\circ}$, which is larger than the one in Fig.~2. \\

The Fe$_6$ chain (Fig.~\ref{fig:S:SymBreaker}b) shows relatively small variation in the lifetime ratios among the atoms. As mentioned before, in an ideal situation, the even-numbered AFM chain should yield $T_{\mathrm H}/T_{\mathrm L}=1$. However, as shown in Fig.~\ref{fig:S:SymBreaker}b, especially at larger $B_x$ values, a finite splitting occurs. We attribute this splitting to slight variations in the $g$-factors of the atoms arising from local imperfections near the atoms, such as defects or edges of the Cu$_2$N island (indicated by white arrows in STM images of Fig.\ref{fig:S:Tilt}). For $B_x < 2$~T, the lifetime ratios measured on each atom are approximately the same, albeit slightly larger than unity. We interpret this as the result of a longitudinal tip field of about $30$~mT causing a similar imbalance on each atom. Around the diabolic point ($3.5$~T), the lifetime ratios approach unity. Meanwhile, at $B_x = 6$~T, the lifetime ratios are about $1.6$ and $0.70$ for even- and odd-numbered atoms in the chain, respectively. This corresponds to an energy difference of about $50$~$\mu$eV, which clearly indicates the Zeeman energy between two ground states induced due to the inhomogeneity among atoms in the chain. \\

\section*{Supplementary Note 6: Investigating $\alpha_{\mathrm{tilt}}$}
\vspace{5mm}
\addcontentsline{toc}{section}{Supplementary Note 6: Investigating $\alpha_{\mathrm{tilt}}$}

While the crystal orientation with respect to the STM stage ($\alpha_{\rm atomic}$) can be estimated by scanning the sample surface at atomic resolution, the alignment between the STM stage and the magnets need to be thoroughly investigated to identify the magnetic fields with respect to the crystal axes. When applying $B_1$, the $B_z$ field expressed in the crystal axes is given by $B_z \approx B_1 \sin(\alpha)$. Due to a slight misalignment ("tilt") between the magnetic field axes and the STM stage, the angle $\alpha_{\rm atomic}$, derived from atomic resolution topographic images, needs to be corrected by a value $\alpha_{\rm tilt}$ such that the angle $\alpha = \alpha_{\rm atomic} - \alpha_{\rm tilt}$. Note that $\alpha_{\rm tilt}$ remains constant throughout our experiments, set by the installation of the STM stage with respect to the magnets, while $\alpha_{\rm atomic}$ varies depending on how the substrate is mounted on the sample holder. In this section, we demonstrate that $\alpha_{\rm tilt} = 3^{\circ}$, estimated from the analysis of lifetime ratios and peak widths of lifetime curves. \\

\begin{figure}[H]
\centering
\includegraphics[width=\linewidth]{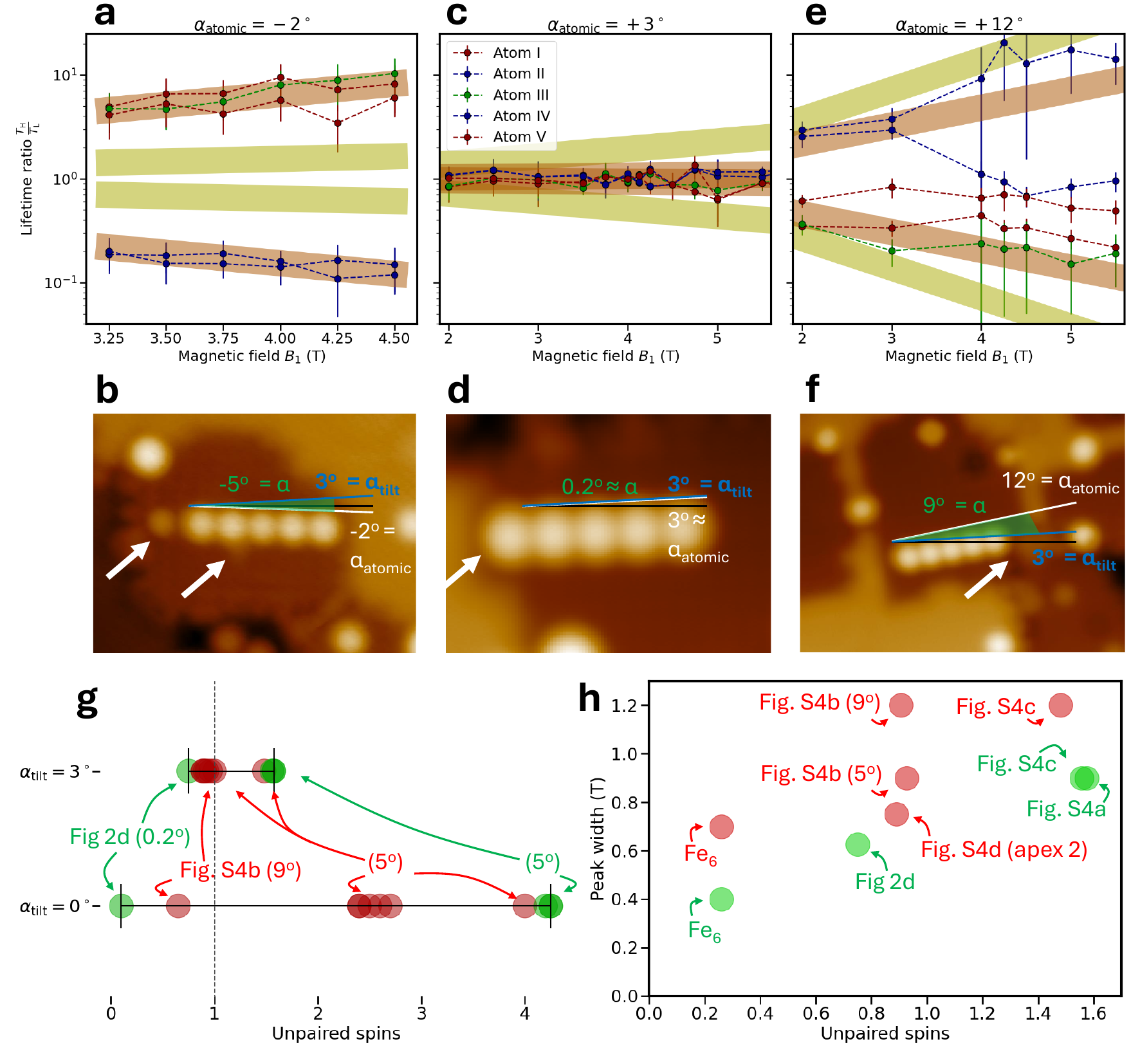}\caption{\textbf{Determining crystal axes.} 
\textbf{(a)} Lifetime ratios $T_{\mathrm H}/T_{\mathrm L}$ and \textbf{(b)} STM image for the Fe$_5$ chain. The corresponding lifetime curve for this chain is given in red in Fig.~\ref{fig:S:Robust}a. The atomic resolution image presents $\alpha_{\rm atomic} = -2^{\circ}$. Considering the Zeeman energy of one ``unpaired spin'' for the Fe$_5$ chain, the calculated lifetime ratios at this angle are represented by the yellow shading, where $\alpha_{\rm tilt} = 0^{\circ}$ and $g$-factors are homogeneous between atoms. However, our measurement shows 4.25 times larger Zeeman splittings, as indicated by the orange shading. \textbf{(c,d)} Same as (a,b), but $\alpha_{\rm atomic} = +3^{\circ}$. The corresponding lifetime data are shown in Fig.~2d. For $\alpha_{\rm atomic} = +3^{\circ}$, the measured Zeeman energy (orange shading) is now 0.1 times smaller than the calculated Zeeman energy (yellow shading). \textbf{(e,f)} Same as (a,b), but $\alpha_{\rm atomic} = +12^{\circ}$. The corresponding lifetime curve is shown in a red curve in Fig.~\ref{fig:S:Robust}b. The lifetime ratios given from the measurement show the Zeeman energy is 0.65 times smaller than the calculated value. Arrows given in each STM image indicate nearby defects that may cause symmetry breaking. The angle $\alpha = \alpha_{\rm tilt} - \alpha_{\rm atomic}$ is given for $\alpha_{\rm tilt} = +3^{\circ}$. \textbf{(g)} Overview of all determined unpaired spins throughout this work before ("$\alpha_{\rm tilt} = 0^{\circ}$") and after ("$\alpha_{\rm tilt} = 3^{\circ}$") adjusting the angle $\alpha_{\rm tilt}$. \textbf{(h)} Overview of unpaired spins and peak width for the various data sets throughout this work. The peak width for atom III is defined as half the range of $B_x$ values over which the lifetime is at most one order of magnitude smaller than the peak lifetime. For panels g and h: Green circles indicate data taken at $V_{\rm bias} = 3$~mV, red circles taken at $5$~mV. Inserted labels indicate the correspondent dataset and the angle $\alpha$.}\label{fig:S:Tilt}
\end{figure}

The peak width of a lifetime curve and the lifetime ratios primarily depend on the energy difference between $\psi_0$ and $\psi_1$, dominated by the Zeeman energy. Accurate determination of the Zeeman energy necessitates the determination of the angle $\alpha$, considering $\alpha_{\mathrm{atomic}}$ and $\alpha_{\mathrm{tilt}}$. In our experiments, crystal orientations can be adjusted by rotating the sample with respect to the sample holder. Figures~\ref{fig:S:Tilt}a--f show lifetime ratios and corresponding STM images for three different chains, each with a different $\alpha_{\rm atomic}$. From STM images, we can extract the angle $\alpha_{\rm atomic}$ for each of the chains: $\alpha_{\mathrm{atomic}} =-2^{\circ}$ for Fig.~\ref{fig:S:Tilt}b, $+3^{\circ}$ for Fig.~\ref{fig:S:Tilt}d, and $+12^{\circ}$ for Fig.~\ref{fig:S:Tilt}f. Depending on $\alpha_{\mathrm{atomic}}$, we observe clear variations in the lifetime ratios and in their deviations between even- and odd-numbered atoms of the chains due to different Zeeman energies. For $\alpha_{\mathrm{atomic}}=3^\circ$ (Fig.~\ref{fig:S:Tilt}c), we observe a negligibly small difference in lifetime ratios between even- and odd-numbered atoms, which already implies the minimal Zeeman energy ($\alpha \approx 0$). Based on the extracted angle $\alpha_{\rm atomic}$, we deduce $B_z$ values relative to applied magnetic fields $B_1$ and, thus, calculate Zeeman energies for each chain. For a single unpaired spin with $S = 2$ for the Fe$_5$ chain, the Zeeman energy is given by $\Delta E \approx 4g\mu_{\mathrm{B}} B_z \approx 4g\mu_{\mathrm{B}} B_1 \sin(\alpha)$. We subsequently calculate lifetime ratios based on the populations of $\psi_0$ and $\psi_1$ through the Boltzmann distribution at the Zeeman energy and $T = 1.8$~K, slightly exceeding the measurement temperature of $T = 1.3\sim1.4$~K (see also Table \ref{tab:SimulationParameters}). \\

Assuming $\alpha_{\rm tilt}=0$ (thus, $\alpha = \alpha_{\mathrm{atomic}}$), the calculated Zeeman energies and corresponding lifetime ratios are depicted by yellow shading in Fig.~\ref{fig:S:Tilt}a,c,e. We found substantial deviations between our measured values and the calculated Zeeman energies for $\alpha_{\rm tilt} = 0$, which suggests the influence of a nonzero $\alpha_{\rm tilt}$ and variations in the $g$-factors among the chain's atoms. As discussed in Supplementary Note 5, we consider the inhomogeneity of $g$-values by assigning a distinct $g$-factor to atom I, differing from the rest ($g_{\rm rest} = 2.11$ following literature \cite{HirjibehedinScience2007}). To optimize $g_1$ and $\alpha_{\rm tilt}$, we initially assume $\alpha_{\rm tilt}=0$ and adjust $g_1$ to scale the calculated Zeeman energies, aligning them with our results. Subsequently, we fine-tune $\alpha_{\mathrm{tilt}}$ to compensate the scaling by $g_1$ to keep $g_1$ within a reasonable range. To characterize both odd- and even-numbered chains uniformly, we introduce the concept of ``{\em unpaired spins}''. For an ideal Fe$_5$ chain ($g_1 = g_{\rm rest}$), there would be one unpaired spin with uncompensated Zeeman energy. Conversely, in an ideal Fe$_6$ chain, no unpaired spins would be present. However, considering inhomogeneity ($g_1 \neq g_{\rm rest}$), the unpaired spin for antiferromagnetically coupled even-numbered chains is defined as ($g_1/g_{\rm rest} - 1$). \\

To match the calculated Zeeman energies with our experimental results, we adjust the values of unpaired spins (i.e., $g_1$). We found optimal agreement when we set the unpaired spins as 4.25, 0.1, and 0.65 for Fig.~\ref{fig:S:Tilt}a,c, and e, respectively, while keeping $\alpha_{\rm tilt}=0$. This is shown in orange shading. Next, we try to compensate for an offset of the angle, $\alpha_{\rm tilt}$, to bring the number of unpaired spins in all measured odd chains as close to unity as possible. This adjustment is carried out as follows: Fig.~\ref{fig:S:Tilt}g shows the determined number of unpaired spins with $\alpha_{\mathrm{tilt}} = 0$ (bottom) and with  $\alpha_{\mathrm{tilt}} = 3^{\circ}$ (top). The overall values of unpaired spins are much closer to 1 for $\alpha_{\mathrm{tilt}} = 3^{\circ}$.
Note that the angles ($\alpha$) used for the simulations in the main text all lie within an error margin of 1 degree compared to the measured angles, after accounting for the tilt ($\alpha_{\mathrm{tilt}}-\alpha_{\mathrm{atomic}}$). Note also that outside of this Supplementary Note $\alpha$ will always be presented as $|\alpha|$. \\

Lastly, using the adjusted tilt angle $\alpha_{\mathrm{tilt}} = 3^{\circ}$ and the corresponding calculated Zeeman energies in terms of unpaired spins, we show the peak width for several different Fe chains as a function of unpaired spins in Fig.~\ref{fig:S:Tilt}h. The green and red symbols indicate data taken at $3$~mV and $5$~mV, respectively. For the latter, we need to consider over-the-barrier transitions, thereby leading to a slight reduction in peak height and an increase in peak width. This effect is clearly observed in the data for Fe$_6$ and the data presented in Fig.~S4c. Furthermore, an increase in Zeeman energy (through an increase in unpaired spins or angle) is also associated with a larger peak width, as expected. Note that the data in Fig.~2d was obtained at a much smaller angle than the rest of the dataset in this overview, resulting in a smaller Zeeman energy. \\

Despite the Fe$_5$ chain having an estimated Zeeman energy of around $30$~$\mu$eV, it is surprising that the peak width in Fig.~2d is somewhat larger than that of the Fe$_6$ chain with a Zeeman energy of about $50$~$\mu$eV. Given the very small energy difference between $\psi_0$ and $\psi_1$ for the data in Fig.~2d, we did not find any convincing indicator to associate the low current in Fig.~2a with N\`eel state A. Without loss of generality we picked this based on Fig.~2b and c, where, for the specific case of $B_x = 4$~T, the average lifetime of the low current state is longer than the average lifetime of the high current state. Code for processing all the lifetime ratios is available in the Open Data folder. \\

\section*{Supplementary Note 7: Lifetime of the Fe$_3$ chain}
\vspace{5mm}
\addcontentsline{toc}{section}{Supplementary Note 7: Lifetime measurements on Fe$_3$}

\begin{figure}[H]
\centering
\includegraphics[width=1\linewidth]{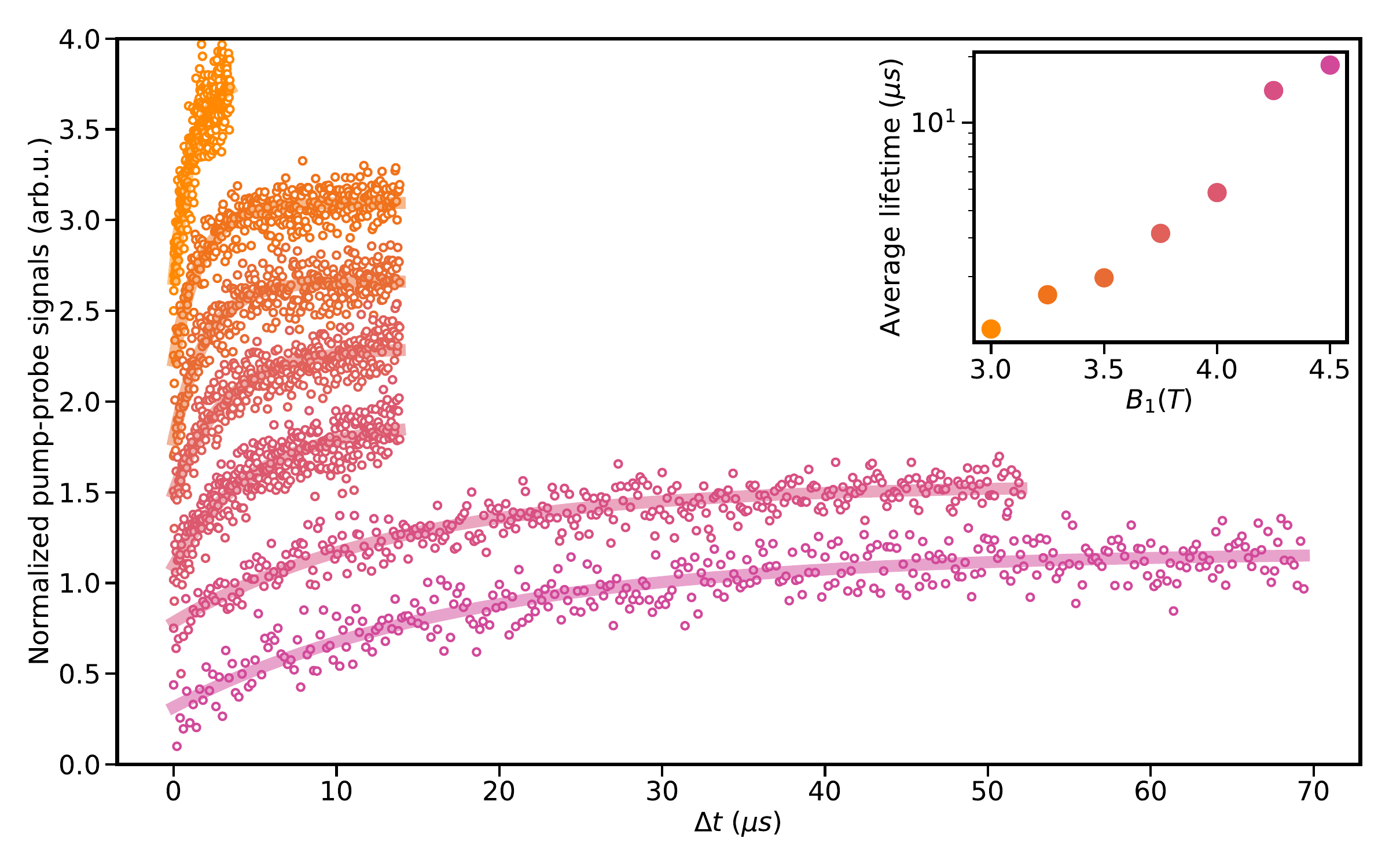}\caption{\textbf{Pump and probe measurements for the Fe$_3$ chain at different $B_x$ fields.} Pump-probe signals measured on the center atom of an Fe$_3$ chain at different $B_x$ magnetic fields. 
Considering the crystal axes with respect to the magnets' axes, increasing the $B_1$ results in increasing both $B_x$ and $B_z$ in a way of $B_z \approx 0.09 \cdot B_x$. With increasing $B_x$, we found the lifetimes monotonically increase (see inset). For the Fe$_3$ chain, our simulation shows the diabolic point (almost) exceeding our magnetic field range. Setpoint: $V_{\rm DC} = 10$~mV, $I = 200$~pA, $V_{\mathrm{pump}} = 35$~mV, $V_{\mathrm{probe}} = 5$~mV, and $T = 1.3$~K. Each pump and probe measurement is offset for clarity.}\label{fig:S5}
\end{figure}

\section*{Supplementary Note 8: Voltage dependence of lifetimes}
\vspace{5mm}
\addcontentsline{toc}{section}{Supplementary Note 8: Voltage dependence}

\begin{figure}[H]
\centering
\includegraphics[width=1\linewidth]{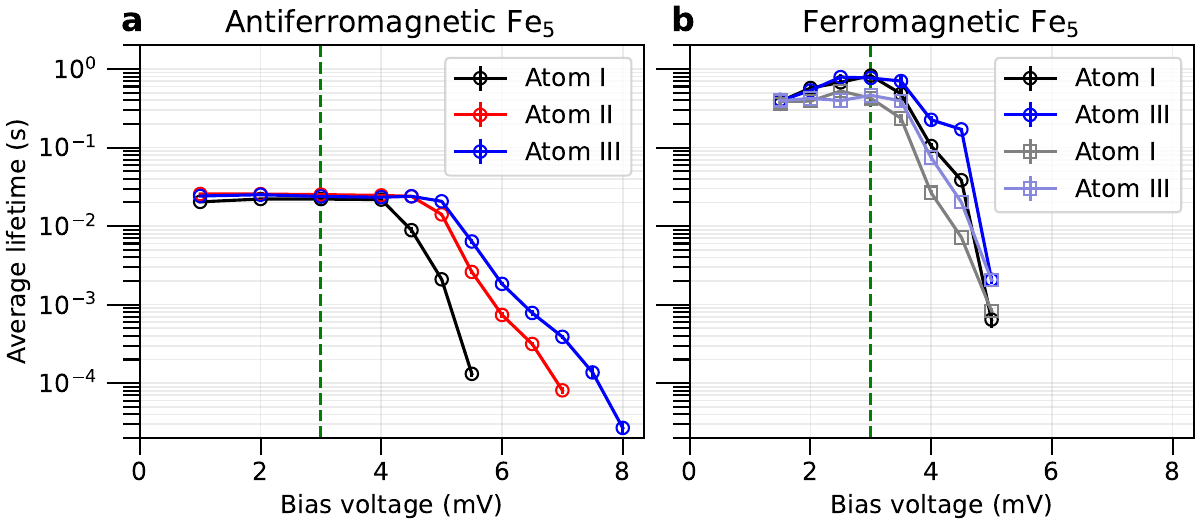}\caption{\textbf{Voltage threshold for over-the-barrier transitions.} \textbf{(a)} Bias dependence of spin lifetimes for the antiferromagnetically coupled Fe$_5$ chain presented in Fig.~2d of the main text measured under constant-current mode at $10\,\mathrm{pA}$. The measurements are shown for three atoms (I--III) in the chain. \textbf{(b)} Bias dependence of spin lifetimes for the ferromagnetically coupled Fe$_5$ chain presented in Fig.~3b. The blue and black circles show the data measured at a constant current of $10\,\mathrm{pA}$. Faint blue and black squares represent the data obtained at a constant tip height (6.6~nS). Lines connecting data points are guides to the eye. The error bars ($2\sigma$) are smaller than the marker size. All data was obtained at $T = 1.4$~K, $B_x = 4.125$~T near the diabolic point for (a) and $B_x = 4.75$~T near the second diabolic point for (b).}\label{fig:biasdep}
\end{figure}

To identify the regimes for the quantum tunneling of magnetization (QTM) and over-the-barrier transitions, we measured the spin lifetimes of Fe$_5$ chains as a function of bias voltages. At bias voltages corresponding to electron energies below the magnetic anisotropy barrier (i.e. less than the energy of $\psi_2$), to first approximation, we expect the lifetimes to be constant, while above the barrier the lifetimes are expected to decrease gradually. Figure~\ref{fig:biasdep}a shows the average spin lifetimes of an antiferromagnetically coupled Fe$_5$ chain. Below $4$~mV the spin lifetime is constant with bias, indicating the QTM regime. Above $4$~mV, spin lifetimes of the outer atom decrease as bias increases. This threshold voltage appears higher for the inner atoms. Due to this variation of threshold voltages between the atoms in the chain, the lifetime curves given in Fig.~3a show strong dependence on the atoms when measured at 5~mV. This observation can be understood by larger values of $J_i$ and $D_i$ further towards the center of the chain, see also Supplementary Note 1. Additionally, owing to the nodal structure of excitation modes, the next lowest excited energy states (e.g. $\psi_2$) are expected to be primarily localized on the outer atoms \cite{SpinelliNatureMaterials2014}. \\

In Fig.~\ref{fig:biasdep}b, we show analogous data for the ferromagnetically coupled Fe$_5$ chain. The blue and black circles present the results measured at the same conditions as in Fig.~\ref{fig:biasdep}a for the central and outer atoms, respectively. Here we clearly observe a maximum lifetime at around $3\,\mathrm{mV}$. This feature likely emerges as a consequence of increased tip-chain exchange interactions at lower bias voltages, which decreases the spin lifetimes. To keep the tip-chain interaction constant, we fixed the conductance to $\sim6.6\,\mathrm{nS}$ and repeated the lifetime measurements at different voltages as given by faint blue and black squares in Fig. \ref{fig:biasdep}b. In this constant height measurement, we found a plateau of the lifetimes below $\sim3.5\,\mathrm{mV}$. Unlike the antiferromagnetic chain, the threshold voltages for the over-the-barrier transitions are similar between the atoms in the chain.
Thus, for both antiferromagnetic and ferromagnetic chains, we chose $3\,\mathrm{mV}$ to characterize the spin lifetimes due to the quantum tunneling of magnetization near the diabolic point. 

\vspace{5mm}

\section*{Supplementary Note 9: Current dependence}
\addcontentsline{toc}{section}{Supplementary Note 9: Current dependence}

\begin{figure}[H]
    \centering
    \includegraphics[scale=0.6]{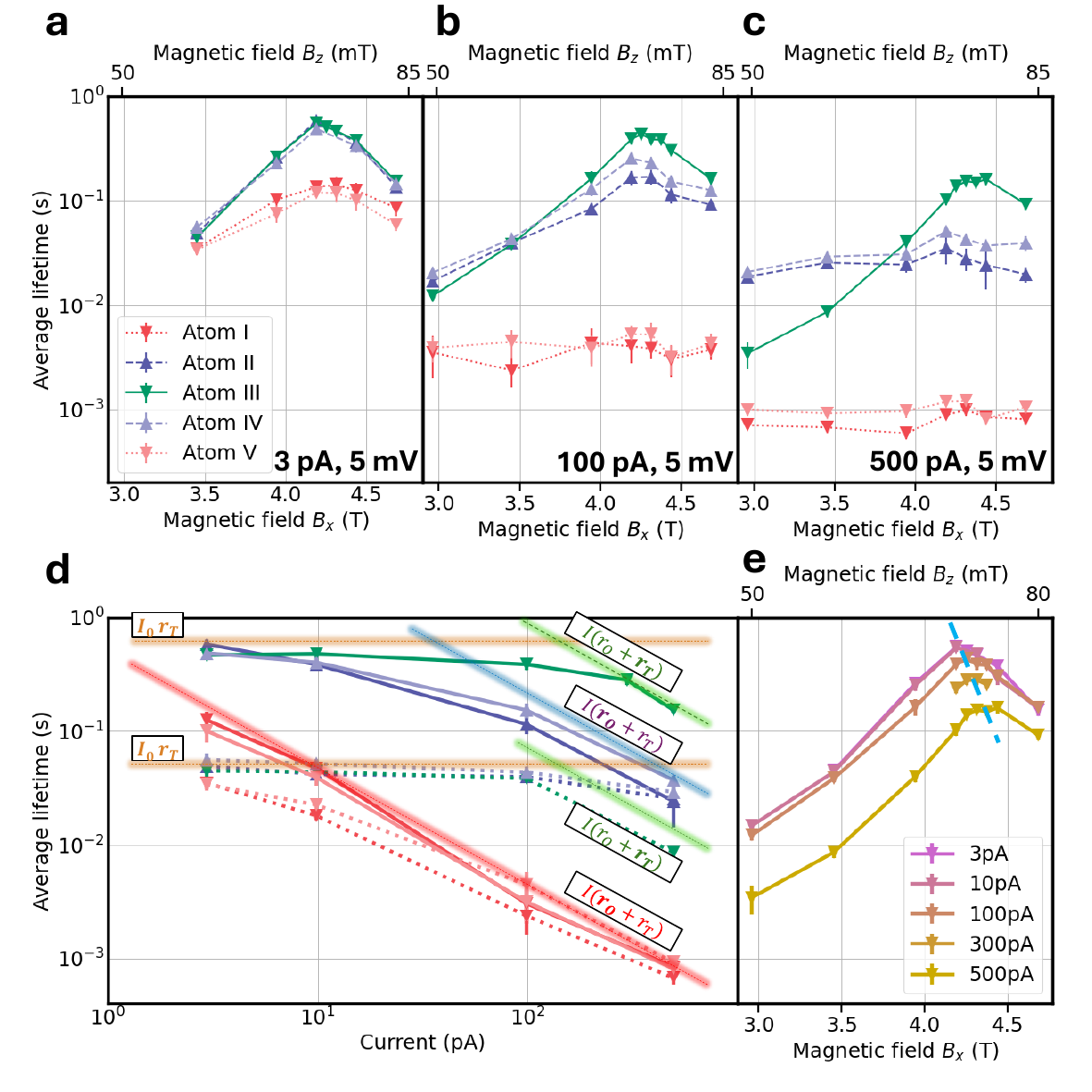}
    \caption{\textbf{Current dependence of spin lifetimes.} \textbf{(a)} Lifetimes measured on each atom at 3~pA, \textbf{(b)} 100~pA, and \textbf{(c)} 500~pA. All data was taken at $5$~mV and $T = 1.3$~K. Associated $B_{\mathrm{z}}$ is shown on the top axes. Lines are guide to the eye. Error bars are $2 \sigma$. \textbf{(d)} Lifetimes of each atom as a function of current at $B_x = 3.5$~T (dotted) and at the first DP $B^{5,1}_{x}$ (solid). Dashed glowing lines represent lifetimes calculated considering transition rates. \textbf{(e)} Lifetimes of atom III as a function of magnetic fields at different currents. The blue dashed line indicates $B^{5,1}_{x}$ used in (d) corresponding to the magnetic fields of the maximum of lifetimes.}
    \label{fig:S:Current}
\end{figure}

In this section, we consider the effects of the current ($I$) on the lifetime of the antiferromagnetic Fe$_5$ chain. Figure~\ref{fig:S:Current} shows the lifetimes measured on each atom in the chain at different currents. Note that the bias voltage was set to $5$~mV, at which lifetime reductions due to over-the-barrier transitions become apparent, most notably on the outer atoms. We thus consider two separate transition rates: $R_{\mathrm{T}}$ for transitions due to tunneling of magnetization and $R_{\mathrm{O}}$ for over-the-barrier transitions. With increasing current, the lifetime generally decreases, due to both $R_{\mathrm{T}}$ and $R_{\mathrm{O}}$ increasing linearly with $I$. The lifetimes can be expressed in terms of transition rates: $(T_{\mathrm{avg}})^{-1} \propto R_{\mathrm{O}} + R_{\mathrm{T}} = Ir_{\mathrm{O}} + (I_0 + I) r_{\mathrm{T}} = I(r_{\mathrm{O}} + r_{\rm T}) + I_0r_{\rm T}$, where the coefficients $r_{\mathrm{O}}$ and $r_{\mathrm{T}}$ are based on the amplitudes of the scattering paths and $I_0$ is the current from the bath interacting with the chain, which has too little energy to induce over-the-barrier excitations. Note that $r_{\mathrm{O}}$ and $r_{\mathrm{T}}$ are independent of $I$ but may be dependent on $V_{\mathrm{bias}}$, $T$, atom location in the chain ($i$), and magnetic fields. We found that $r_{\mathrm{O}}$ does not depend on the transverse magnetic field, but $r_{\mathrm{T}}$ does.  \\

Figures~\ref{fig:S:Current}a--c show the lifetimes measured for each atom in the Fe$_5$ chain at a current of 3~pA, 100~pA and 500~pA, respectively. When increasing the current, the outer atoms show no signature of a diabolic point, as $R_{\rm O}$ becomes too large. In contrast, the diabolic point remains visible for atom III but appears at higher magnetic fields with overall reduced lifetimes (Fig.~\ref{fig:S:Current}e). The dashed blue line in Fig.~\ref{fig:S:Current}e indicates the shift of the DP as a result of the tip field. Increasing the current leads to smaller tip-sample distances and larger tip fields, resulting in the DP at larger external magnetic fields. This indicates that the tip field is counteracting the external magnetic field. Note that this corresponds to tip 2 of Fig.~\ref{fig:S:Robust}d. Importantly, atom III, not limited by over-the-barrier transitions, shows that the tip field does not destroy the appearance of the diabolic point, despite the field being applied to only one atom of the chain. This further supports the claim that this effect is robust to many variations in the parameters of the Hamiltonian.\\

This current dependence is further analyzed in Fig.~\ref{fig:S:Current}d by plotting the lifetimes as a function of current at each magnetic field of $B_x = 3.5$~T (dotted lines) and $B_x = B^{5,1}_{x}$ (solid lines). We found four different effects of the current on the lifetimes:

\begin{enumerate}
    \item As indicated by the orange lines in Fig.~\ref{fig:S:Current}d, for the inner atoms, the lifetimes are nearly constant for smaller $I$, where the plateau value and the threshold current depend on $B_x$. At the plateaus, the lifetimes are mostly determined by $I_0 r_{\rm T}$, which depends on the scattering intensity and, thus, on $B_x$. Consistent with the rest of this work, in the absence of over-the-barrier transitions, a longer lifetime is observed at $B_x$ closer to the DP.
    \item Excitation over the barrier due to an applied bias and finite temperature causes $R_{\mathrm{O}}$ to limit the lifetime for all atoms \cite{PaulNPhysics2017}. At $V_{\mathrm{bias}} = 5$~mV and $T = 1.3$~K, this effect is less significant towards the center of the chain (see Supplementary Note 8). This effect does not depend on field and should therefore result in constant lifetimes throughout the full range of $B_x$. This is the case for atoms I/V in Figs.~\ref{fig:S:Current}b,c and atoms II/IV in Fig.~\ref{fig:S:Current}c. This contribution scales linearly with current, as depicted by the red and blue glowing straight lines in Fig.~\ref{fig:S:Current}d. \\
    \item Additional electrons might cause through-the-barrier transitions if the electrons do not have enough energy to cause over-the-barrier transitions. We find that for the experimental conditions for Fig.~\ref{fig:S:Current}, only atom III shows this behavior, as the shape of the lifetime curves in Figs.~\ref{fig:S:Current}a--c does not really flatten with increasing current. These transition events also depend on the scattering amplitude between the lowest energy eigenstates. In Fig.~\ref{fig:S:Current}d, this is highlighted with the green glowing dashed lines, which intersects with the orange lines at around 200~pA. This suggests the rate of bath electrons interacting with the system is of a similar order.    \\
    \item In Fig.~\ref{fig:S:Current}c, the lifetimes of atoms II/IV appear higher than atom III for $B<3.75$~T, which can be attributed to the atomic exchange bias as the magnetic tip approaches close to the chain at higher current. This exchange bias can be modeled as an increased $B_z$ for atoms II/IV and a decreased $B_z$ for atoms I/III/V in their ground state \cite{YanNature2014}. For $B > 3.75$~T, the lifetimes of atoms II/IV are limited by over-the-barrier transitions, while the atom III is free from this and, thus, shows longer lifetime. 

\end{enumerate}

\section*{Supplementary Note 10: Methods and Data Acquisition}
\vspace{5mm}
\addcontentsline{toc}{section}{Supplementary Note 10: Methods and Data Acquisition}

\subsection*{Sample Preparation}
We used a home-built STM system~\cite{HwangRSI2022}, operating at $1.3$~K and $B = 0-6$~T in the plane of the sample, mainly perpendicular to the axis of the chain. The Cu$_2$N/Cu(100) sample was prepared as described in \cite{HirjibehedinScience2006}. The tip was prepared as in \cite{LothScience2012}. Fe atoms on the Cu$_2$N were picked up by applying voltage pulses of $\sim 1$~V (setpoint of $100$~pA, $20$~mV, then moved $-300$~pm), and dropped at $-600$~$\mu$V with the tip gradually approaching the surface until an abrupt change in current was observed. The Fe was subsequently hopped into place with a pulse of $\sim 1$~V at ($100$~pA, $20$~mV). Preferred hopping directions were determined by straining of the Cu$_2$N lattice in line with previous works \cite{YamadaJSSN2023}, and utilized for efficient construction of the chains. 

\subsection*{Data Acquisition}
The spin-polarized STM tip was prepared by attaching several Fe atoms to the Cu-coated tip apex. Using this spin-polarized tip, the magnetization switching was measured in either a constant-current or constant-height mode. To measure the switching in the order of a millisecond or below, a DAQ (NI 782258-01) was used to record an incoming current stream of up to 10 seconds with the feedback turned off with a sampling rate of $10$~kHz. For slightly longer lifetimes, the data was recorded directly through the internal DAQ of Nanonis electronics with a sampling rate of $2$~kHz. For lifetimes longer than $100$~ms, the feedback was turned on with an extremely long time constant in the feedback loop. This allows the feedback to account for drift, but keep the switching signals in the current data stream. For very long lifetimes ($\geq 1$~s), a constant-current mode was used such that the variation of tip heights was used to determine the magnetization switches. The tip's feedback was set such that the response time is much faster than the average switch time. The gradual drift of the tip height was subtracted from the data. The angle $\alpha$ was derived from the angle determined by atomic resolution STM images, and adjusted to include a tilt of $3^{\circ}$, see Supplementary Note 6.

\vspace{5mm}

\bibliographystyle{apsrev4-2}
\bibliography{references}

\end{document}